\documentclass[12pt]{article}
\usepackage{a4wide,epsfig,amsmath,amssymb,cite,scalefnt,graphicx}
\numberwithin{equation}{section}
\allowdisplaybreaks 
\usepackage{array}
\usepackage{epstopdf}
\usepackage{tikz}
\usetikzlibrary{intersections}
\usetikzlibrary{calc}
\usetikzlibrary{shapes}
\usetikzlibrary{arrows,shapes,positioning}
\usetikzlibrary{decorations.markings}
\tikzstyle arrowstyle=[scale=1]
\tikzstyle directed=[postaction={decorate,decoration={markings,
    mark=at position .5 with {\arrow[arrowstyle]{stealth}}}}]
\tikzstyle reverse directed=[postaction={decorate,decoration={markings,
    mark=at position .5 with {\arrowreversed[arrowstyle]{stealth};}}}]
\newcommand{\vertex}{\node[fill,circle,inner sep=0pt,minimum size=0pt]}
\newcommand{\coord}[1]{({sin(#1)},{cos(#1)})}

\def\be{\begin{equation}}
\def\ee{\end{equation}}
\def\nn{\nonumber\\}

\def\Qbar{\overline Q}
\def\cbar{\overline c}

\def\frakk[#1#2{{{#1}\over{#2}}}

\def\Ytil{\tilde Y}

\def\atil{\tilde \alpha}
\def\ctil{\tilde c}
\def\dtil{\tilde \delta}

\def\Ytil{\tilde Y}

\def\pa{\partial}

\input amssym.def
\input amssym
\def\pa{\partial}
\def\be{\begin{equation}}
\def\ee{\end{equation}}
\def\nn{\nonumber\\}

\def\Qbar{\overline Q}

\def\Ctil{{\tilde C}}

\def\tkappa{{\tilde\kappa}}

\def\Khat{\hat K}
\def\Rbar{\overline {R}}
\def\Jtil{\tilde J}
\def\tbar{\overline t}
\def\atil{\tilde a}
\def\abar{\overline a}
\def\ctil{\tilde c}
\def\cbar{\overline c}
\def\dtil{\tilde d}
\def\dbar{\overline d}
\def\cirk{\,{\raise1pt \hbox{${\scriptscriptstyle \circ}$}}\,}

\def \olr{{\raise6.5pt\hbox{$\leftrightarrow  \! \! \! \! \!$}}}

\newcommand{\vbeta}{\node[fill=black,diamond,inner sep=3.5pt,minimum size=0pt]}
\newcommand{\vsq}{\node[fill=black,rectangle,inner sep=3.5pt,minimum size=0pt]}
\newcommand{\vcirc}{\node[fill=black,circle,inner sep=2pt,minimum size=0pt]}

\font\ninerm=cmr9 \font\ninesy=cmsy9
\font\eightrm=cmr8 \font\sixrm=cmr6
\font\eighti=cmmi8 \font\sixi=cmmi6
\font\eightsy=cmsy8 \font\sixsy=cmsy6
\font\eightbf=cmbx8 \font\sixbf=cmbx6
\font\eightit=cmti8
\def\eightpoint{\def\rm{\fam0\eightrm}
  \textfont0=\eightrm \scriptfont0=\sixrm \scriptscriptfont0=\fiverm
  \textfont1=\eighti  \scriptfont1=\sixi  \scriptscriptfont1=\fivei
  \textfont2=\eightsy \scriptfont2=\sixsy \scriptscriptfont2=\fivesy
  \textfont3=\tenex   \scriptfont3=\tenex \scriptscriptfont3=\tenex
  \textfont\itfam=\eightit  \def\it{\fam\itfam\eightit}%
  \textfont\bffam=\eightbf  \scriptfont\bffam=\sixbf
  \scriptscriptfont\bffam=\fivebf  \def\bf{\fam\bffam\eightbf}%
  \normalbaselineskip=9pt
  \setbox\strutbox=\hbox{\vrule height7pt depth2pt width0pt}%
  \let\big=\eightbig  \normalbaselines\rm}
\catcode`@=11 %
\def\eightbig#1{{\hbox{$\textfont0=\ninerm\textfont2=\ninesy
  \left#1\vbox to6.5pt{}\right.\n@@space$}}}
\def\vfootnote#1{\insert\footins\bgroup\eightpoint
  \interlinepenalty=\interfootnotelinepenalty
  \splittopskip=\ht\strutbox %
  \splitmaxdepth=\dp\strutbox %
  \leftskip=0pt \rightskip=0pt \spaceskip=0pt \xspaceskip=0pt
  \textindent{#1}\footstrut\futurelet\next\fo@t}
\catcode`@=12 %
\def\today{\number\day\ \ifcase\month\or January\or February\or March\or
April\or May\or June\or July\or
August\or September\or October\or November\or December\fi, \number\year}

\input epsf

\begin{document}
\numberwithin{equation}{section}

\begin{titlepage}
\begin{flushright}
LTH1239\\

\end{flushright}
\date{}
\vspace*{3mm}

\begin{center}
{\Huge Anomalous dimensions for $\phi^n$ in scale invariant $d=3$ theory}\\[12mm]
{\bf I.~Jack\footnote{\tt dij@liverpool.ac.uk} and D.R.T.~Jones\footnote{\tt drtj@liverpool.ac.uk}
}\\

\vspace{5mm}
Dept. of Mathematical Sciences,
University of Liverpool, Liverpool L69 3BX, UK\\

\end{center}

\vspace{3mm}
\begin{abstract}
Recently it was shown that the scaling dimension of the operator $\phi^n$ in scale-invariant $d=3$ theory may be computed semiclassically, and this was verified to leading order (two loops) in perturbation theory at leading and subleading $n$. Here we extend this verification to six loops, once again at leading and subleading $n$. We then perform a similar exercise for a theory with a multiplet of real scalars and an $O(N)$ invariant hexic interaction. We also investigate the strong-coupling regime for this example.
\end{abstract}

\vfill

\end{titlepage}

\section{Introduction}

Renormalizable theories with scale invariant scalar self-interactions exist in four ($\phi^4$), six $(\phi^3)$ and three ($\phi^6$) dimensions. There has been considerable recent interest in the latter, in particular in theories involving complex scalar fields and a $U(1)$ invariance, with $(\phi^*\phi)^3$-type interactions\footnote{In Ref.~\cite{Jack:2015tka}, a $d=3$ Chern-Simons gauge theory with such an interaction was studied, including also Yukawa couplings to  a multiplet of fermions. From the two-loop $\beta$ functions (previously calculated by Avdeev et al\cite{Av1}\cite{Av2}), persuasive evidence was presented for the existence of an $a$-theorem for such theories. This was subsequently confirmed by higher loop calculations in Ref.~\cite{Jack:2016utw}, with the scalar self-interaction replaced by a general hexic one with an arbitrary number of real scalar fields.}. The anomalous dimension of the operator $\phi^n$, $\gamma_{\phi^n}$,  was calculated at the two loop level (in usual perturbation theory in powers of coupling constant $\lambda)$ in Ref.~\cite{Bad} for the $U(1)$ invariant pure scalar theory, and the result compared with a semiclassical calculation valid to all orders in the product $\lambda n$; following the lines of similar calculations in four dimensions\cite{Bad2}. Of particular interest in this context is large $n$, because large charge operators are of peculiar 
relevance in conformal field theory. Also, amplitudes corresponding to many external lines are increasingly relevant in particle physics phenomenology, as collider energies increase, so insights gained by the study of them in simpler theories is worthwhile in itself. Agreement was found in Ref.~\cite{Bad} between perturbative and semiclassical results at the level of the leading and sub-leading terms in an expansion in powers of $1/n$.  Here we extend the straightforward perturbative calculation to the six loop level and once again obtain agreement with the semiclassical calculation for the two leading terms in the same expansion. 
 
In Ref.~\cite{sann}, the calculations of Ref.~\cite{Bad} were extended from the $U(1)$ to the $O(N)$ case. Accordingly, we perform our perturbative checks up to six loops for the semiclassical $O(N)$ result as well. Furthermore, the $U(1)$ semiclassical result was compared with an effective field theory valid at large $\lambda n$. Accordingly we also examine the semiclassical $O(N)$ results at large charge and find that we can obtain exact results for the $N$-dependent part of the coefficients in a large-charge expansion.

The paper is organised as follows: In Section 2 we describe the semiclassical calculation in the $U(1)$ case, following Ref.~\cite{Bad}, and then compare with the perturbative(i.e. small $\lambda n$) results at four and six loops. In Section 3 we discuss the extension to the $O(N)$ case as in Ref.~\cite{sann}, and perform a similar perturbative comparison. In Section 4 we describe the large charge limit and show how to compute the $N$-dependent parts of the coefficients in the large charge expansion. We offer some concluding remarks in Section  5. Finally in the Appendix we give a pedagogical description of the various methods used in our computation of the four-loop and six-loop Feynman diagrams involved in our perturbative check.

\section{The $U(1)$ case}

The lagrangian of the theory is
\be 
{\cal L}=\pa_{\mu}\phi^*\pa^{\mu}\phi+m^2\phi^*\phi+\left(\frac{\lambda}{3!}\right)^2(\phi^*\phi)^3
\ee
We shall be using dimensional regularisation with $d=3-\epsilon$. The agreement between the semiclassical and perturbative results is expected to hold at the conformally invariant fixed point. However, because the $\beta$-function starts at two-loop order in $d=3$, the theory is conformally invariant up to ${\cal O}(\lambda)$, and this is already sufficient for the agreement of the leading and subleading terms in $n$.
The scaling dimension $\Delta_{\phi^n}$ is expanded as (returning to general $d$ for the present, in order to facilitate the later discussion of convergence issues)
\be
\Delta_{\phi^n}=n\left(\frac d2-1\right)+\gamma_{\phi^n}=\sum_{\kappa=-1}\lambda^{\kappa}\Delta_{\kappa}(\lambda n).
\label{scal}
\ee
For the leading and subleading terms in $n$, knowledge of $\Delta_{-1}$ and $\Delta_0$ is sufficient. The semiclassical computation is performed by mapping the theory via a Weyl transformation to a cylinder $\mathbb{R}\times S^{d-1}$, where $S^{d-1}$ is a sphere of radius $R$; where the ${\cal R}\phi^*\phi$ term (${\cal R}$ being the Ricci curvature) generates an effective $m^2\phi^*\phi$ mass term with $m=\frac{d-2}{2R}$. It was shown in Ref.~\cite{Bad} that  stationary configurations of the action are characterised by a chemical potential $\mu$, 
\be
R\mu=\frac{1}{2\sqrt2}\sqrt{1+\sqrt{1+\frac{\lambda^2n^2}{12\pi^2}}}.
\label{mudef}
\ee
It was further shown in Ref.~\cite{Bad} that $\Delta_{-1}$ may be written
\be
\Delta_{-1}(\lambda n)=\lambda nF_{-1}\left(\frac{\lambda^2n^2}{12\pi^2}\right),
\label{Delmin}
\ee
where
\be
F_{-1}(x)=\frac{1+ \sqrt{1+x}+\tfrac x3}{\sqrt2(1+\sqrt{1+x})^{\frac32}}.
\label{Fdef}
\ee
(For convenience we give the results for $d=3$ in Eqs.~\eqref{mudef}, \eqref{Fdef}.)
Expanding to quadratic order around stationary configurations results in an action with two modes $\omega_{\pm}$ given by
\be
\omega_{\pm}^2(l)=J_l^2+2(2\mu^2-m^2)\pm2\sqrt{J_l^2\mu^2+(2\mu^2-m^2)^2},
\label{omdef}
\ee
where
\be
J_l^2=\frac{l(l+d-2)}{R^2}
\label{Jdef}
\ee
is the eigenvalue of the Laplacian on the sphere. The dispersion relation for $\omega_+$ describes a ``gapped'' mode, while that for $\omega_-$ describes a ``Type I'' (relativistic) Goldstone boson\cite{Niel}. The one-loop correction $\Delta_0$ is then determined by the fluctuation determinant corresponding to this quadratic action, which is given by
\be
\Delta_0(\lambda n)=\frac{R}{2}\sum_{l=0}^{\infty}n_l\left[\omega_+(l)+\omega_-(l)-2\omega_0(l)\right],
\label{delzero}
\ee
where 
\be
\omega_0^2(l)=J_l^2+m^2=\frac{1}{R^2}\left(l+\frac{d-2}{2}\right)^2
\ee
is the free theory dispersion relation,
\be
n_l=\frac{(2l+d-2)\Gamma(l+d-2)}{\Gamma(l+1)\Gamma(d-1)}
\label{nldef}
\ee
 is the multiplicity of the laplacian on the $d$-dimensional sphere, and where $\omega_{\pm}$ are defined in \eqref{omdef}. It was shown in Ref.~\cite{Bad} that after analytic continuation to negative $d$, we may obtain a regularised form for $\Delta_{0}$  convergent for $d\rightarrow3$, and we obtain in $d=3$
\be
\Delta_{0}(\lambda n)=\frac14-3(R\mu)^2+\tfrac12\sqrt{8R^2\mu^2-1}+\frac12\sum_{l=1}^{\infty}\sigma(l),
\label{delreg}
\ee
where $\mu$, $\omega_{\pm}$ are given by Eqs.~\eqref{mudef}, \eqref{omdef} (with now $m=\frac{1}{2R}$) and where
\be
\sigma(l)=(1+2l)R[\omega_+(l)+\omega_-(l)]-4l(l+1)-\left(6R^2\mu^2-\frac12\right)
\label{sigdef}
\ee
is defined by subtracting positive and zero powers of $l$ in the large-$l$ expansion of Eq.~\eqref{delzero} so as to give a convergent sum. With a slight abuse of notation, we use the same notation $\Delta_0(\lambda n)$ for both the unregularised and regularised forms of the fluctuation operator.

\begin{figure}[ht]
\begin{tikzpicture}
\matrix[column sep = 1cm]
{
	\node (vert_cent) {\hspace{-13pt}$\phantom{-}$};
	\vertex at \coord{-90} (A) {};
	\vertex at \coord{150} (B) {};
	\vertex at \coord{30} (C) {};
	\draw  [directed,bend left=70] (C) to (B);
         \draw  [directed,bend left=70] (A) to (C);
           \draw  [directed,bend left=-75] (A) to (B) ;
           \draw  [directed,bend left=20] (A) to (C) ;
           \draw  [directed,bend left=30] (A) to (B) ;
           \draw  [directed,bend left=-25] (A) to (C) ;
\draw [directed] (C) to (0.616,1.367);
\draw [directed] (C) to (0.876,1.218);
\draw [directed] (B) to (0.477,-1.422);
\draw [directed] (B) to (0.993,-1.124);
 \draw [directed] (B) to (0.75,-1.3);

\vbeta at (A) {};
\draw (0,-1.5) node {\small{(A)}};
	
&
	\node (vert_cent) {\hspace{-13pt}$\phantom{-}$};
	\vertex at \coord{-90} (A) {};
	\vertex at \coord{150} (B) {};
	\vertex at \coord{30} (C) {};
          \draw [directed,bend left=-30] (C) to (B);
	\draw  [reverse directed,bend left=70] (C) to (B);
         \draw  [directed,bend left=70] (A) to (C);
           \draw  [directed,bend left=-75] (A) to (B) ;
           \draw  [directed,bend left=30] (A) to (B) ;
           \draw  [directed,bend left=-25] (A) to (C) ;
\draw [directed] (C) to (0.616,1.367);
\draw [directed] (C) to (0.876,1.218);
\draw [directed] (B) to (0.616,-1.367);
\draw [directed] (B) to (0.876,-1.218);

\vbeta at (A) {};
\draw (0,-1.5) node {\small{(B)}};

&
	\node (vert_cent) {\hspace{-13pt}$\phantom{-}$};
	\vertex at \coord{-90} (A) {};
	\vertex at \coord{150} (B) {};
	\vertex at \coord{30} (C) {};
	\draw  [directed,bend left=70] (C) to (B);
          \draw [directed,bend left=-30] (C) to (B);
         \draw  [directed,bend left=70] (A) to (C);
           \draw  [directed,bend left=-75] (A) to (B) ;
           \draw  [directed,bend left=20] (A) to (C) ;
           \draw  [directed,bend left=-25] (A) to (C) ;
\draw [directed] (C) to (0.75,1.3);
\draw [directed] (B) to (0.477,-1.422);
\draw [directed] (B) to (0.993,-1.124);
 \draw [directed] (B) to (0.75,-1.3);

\vbeta at (A) {};
	;
\draw (0,-1.5) node {\small{(C)}};

&
	\node (vert_cent) {\hspace{-13pt}$\phantom{-}$};
	\vertex at \coord{-90} (A) {};
	\vertex at \coord{150} (B) {};
	\vertex at \coord{30} (C) {};
          \draw [directed,bend left=10] (C) to (B);
	\draw  [reverse directed,bend left=-30] (C) to (B);
         \draw  [directed,bend left=70] (A) to (C);
           \draw  [directed,bend left=-75] (A) to (B) ;
           \draw  [directed,bend left=-25] (A) to (C) ;
            \draw  [directed,bend left=75] (C) to (B) ;
  \draw [directed] (C) to (0.75,1.3);
\draw [directed] (B) to (0.616,-1.367);
\draw [directed] (B) to (0.876,-1.218);

\vbeta at (A) {};
\draw (0,-1.5) node {\small{(D)}};

&
	\node (vert_cent) {\hspace{-13pt}$\phantom{-}$};
	\vertex at \coord{-90} (A) {};
	\vertex at \coord{150} (B) {};
	\vertex at \coord{30} (C) {};
          \draw [directed,bend left=10] (C) to (B);
          \draw [reverse directed,bend left=-90] (C) to (B);
	\draw  [reverse directed,bend left=-30] (C) to (B);
         \draw  [directed,bend left=70] (A) to (C);
           \draw  [directed,bend left=-75] (A) to (B) ;
            \draw  [directed,bend left=75] (C) to (B) ;
  \draw [directed] (C) to (0.75,1.3);
 \draw [directed] (B) to (0.75,-1.3);
\vbeta at (A) {};
\draw (0,-1.5) node {\small{(E)}};
	
\\
};
\end{tikzpicture}
\caption{Four loop diagrams corresponding to $\gamma_{\phi^n}$} \label{diagfour}
\end{figure}
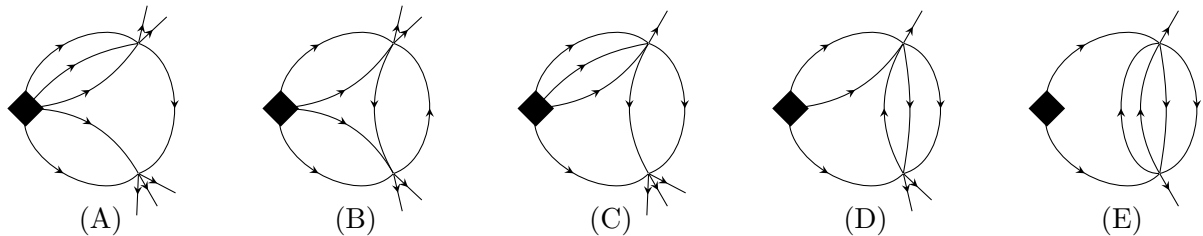

Expanding $\sigma(l)$ in powers of $\frac{\lambda^2n^2}{12\pi^2}$, 
\be
\sigma(l)=C_{2,l}\left(\frac{\lambda^2n^2}{12\pi^2}\right)^2+C_{3,l}\left(\frac{\lambda^2n^2}{12\pi^2}\right)^3+\ldots,
\label{sigexp}
\ee
where
\begin{align}
C_{2,l}=&-\frac{13l^2+13l+1}{128l(l+1)(2l+1)^2},\nn
C_{3,l}=&\frac{208l^6+624l^5+747l^4+454l^3+138l^2+15l+1}{1024l^2(l+1)^2(2l+1)^4}.
\label{Cdef}
\end{align}
Using
\begin{align}
\sum_{l=1}^{\infty}C_{2,l}=&\frac{1}{16}\left(1-\frac{9}{64}\pi^2\right),\nn
\sum_{l=1}^{\infty}C_{3,l}=&\frac{1}{1024}\left(-47+\frac{10}{3}\pi^2+\frac{9}{32}\pi^4\right),
\label{Csum}
\end{align}
and combining Eqs.~ \eqref{scal} (with now $d=3$), \eqref{Delmin}, \eqref{delreg}-\eqref{Csum} we find the expansion 
\begin{align}
\Delta_{\phi^n}=&\frac n2 +\kappa\left[\frac{n^3-3n^2}{9}+{\cal O}(n)\right]\nn
&-\kappa^2\left[\frac{n^5}{9}-\frac{n^4(64-9\pi^2)}{72}+{\cal O}(n^3)\right]\nn
&+\kappa^3\left[\frac{2n^7}{9}+\frac29\left\{-13+\frac{10}{9}\pi^2+\frac{3}{32}\pi^4\right\}n^6+{\cal O}(n^5)\right]+\ldots,
\label{Delexp}
\end{align}
where $\kappa = \left(\frac{\lambda}{8\pi}\right)^2$.
The leading $n$ term in $\kappa$ was first given in Ref.~\cite{rod}.
The terms up to ${\cal O}(\kappa^2)$ were given in Ref.~\cite{Bad}, where the agreement with perturbative calculations was also checked at two loops (${\cal O}(\kappa)$).
We shall now continue the perturbative check of the semiclassical results to four and then six loops.
The four-loop diagrams are shown in Fig.~\ref{diagfour}. The lozenge represents the location of the $\phi^n$ vertex. The extraction of the poles in $\epsilon$ from these diagrams is described in some detail in the Appendix.
The results for the various diagrams in terms of a basic set of Feynman integrals is shown in  Table~\ref{4loop} (a factor of $\kappa^2$ is also implicitly assumed for each graph).
\begin{table}[ht]
\begin{center}
\begin{tabular}{|c|c|} \hline
&\\
Graph& Result\\
&\\
\hline
&\\
$A$ & $\frac{n(n-1)(n-2)(n-3)(n-4)}{144}I_4$\\
&\\
\hline
& \\
$B$ & $\frac{n(n-1)(n-2)(n-3)}{32}I_{4bbb}$\\
&\\
\hline
&\\
$C$ & $\frac{n(n-1)(n-2)(n-3)}{72}I_4$\\
&\\
\hline
&\\
$D$ & $\frac{n(n-1)(n-2)}{8}I_4$\\
&\\
\hline
&\\
$E$ & $\frac{n(n-1)}{8}\Ytil$\\

&\\ 
\hline
\end{tabular}
\caption{\label{4loop}Four-loop results for contributions to $\gamma_{\phi^n}$}
\end{center}
\end{table}
The notation for these integrals conforms to Ref.~\cite{Gracey:2016tuh}. Using the results for these integrals as  listed in that paper, and also recapitulated in the Appendix, the total for 
the four loop anomalous dimension is thus (remembering to multiply the simple pole contribution by a loop factor of four)
\footnote{Eq.~\eqref{gamfour} represents the contribution from the graphs shown in Fig.~\ref{4loop}. We have omitted contributions from graphs with self energy insertions on the external lines, which, however, contribute only to the term linear in $n$.}
\be
\gamma_{\phi^n}^{(4)} 
= -\kappa^2 \frac{n(n-1)}{72}\left[9\pi^2 (n-2)(n-3)+8n^3-56n^2+272n-456\right].
\label{gamfour}
\ee
This result can in principle also be extracted from expressions derived in Refs.~\cite{hughod} and \cite{hughster}.
 Expanding in powers of $1/n$ we obtain
\be
\gamma_{\phi^n}^{(4)} 
= -\frac{\kappa^2}{9}\left[n^5 + n^4\left(\frac{9\pi^2}{8}-8\right)  +\cdots\right], 
\ee
in agreement with Eq.~\eqref{Delexp} and Eq.~(31) of \cite{Bad}. 

\begin{figure}[ht]
\center\begin{tikzpicture}
\matrix[column sep = 1cm]
{

	\node (vert_cent) {\hspace{-13pt}$\phantom{-}$};
	\vertex at \coord{-90} (A) {};
	\vertex at \coord{150} (B) {};
	\vertex at \coord{30} (C) {};
	\vertex at \coord{90} (D) {};
           \draw [directed] (A) to (D);
	\draw  [directed,bend left=0] (B) to (D);
	\draw  [directed,bend left=0] (C) to (D);
           \draw  [directed,bend left=-70] (A) to (B) ;
           \draw  [directed,bend left=-20] (A) to (B) ;
           \draw  [directed,bend left=25] (A) to (B) ;
           \draw  [directed,bend left=20] (A) to (C) ;
           \draw  [directed,bend left=-25] (A) to (C) ;
         \draw  [directed,bend left=70] (A) to (C);
\draw [directed] (C) to (0.616,1.367);
\draw [directed] (C) to (0.876,1.218);
\draw [directed] (B) to (0.616,-1.367);
\draw [directed] (B) to (0.876,-1.218);
\draw [directed] (D) to (1.35,-0.35);
\draw [directed] (D) to (1.35,0.35);
\draw [directed] (D) to (1.5,0);

\vbeta at (A) {};
\vcirc at (D) {};
	;
\draw (0,-1.5) node {\small{(a)}};

&

	\node (vert_cent) {\hspace{-13pt}$\phantom{-}$};
	\vertex at \coord{-90} (A) {};
	\vertex at \coord{150} (B) {};
	\vertex at \coord{30} (C) {};
	\vertex at \coord{90} (D) {};
           \draw [directed,bend left=10] (A) to (D);
           \draw [directed,bend left=-10] (A) to (D);
	\draw  [directed,bend left=0] (D) to (B);
	\draw  [directed,bend left=0] (C) to (D);
           \draw  [directed,bend left=-70] (A) to (B) ;
           \draw  [directed,bend left=10] (A) to (B) ;
           \draw  [directed,bend left=-15] (A) to (C) ;
           \draw  [directed,bend left=20] (A) to (C) ;
         \draw  [directed,bend left=70] (A) to (C);
\draw [directed] (C) to (0.616,1.367);
\draw [directed] (C) to (0.876,1.218);
\draw [directed] (D) to (1.43,-0.25);
\draw [directed] (D) to (1.43,0.25);
\draw [directed] (B) to (0.477,-1.422);
\draw [directed] (B) to (0.993,-1.124);
 \draw [directed] (B) to (0.75,-1.3);

\vbeta at (A) {};
\vcirc at (B) {};
	;
\draw (0,-1.5) node {\small{(b)}};

&

	\node (vert_cent) {\hspace{-13pt}$\phantom{-}$};
	\vertex at \coord{-90} (A) {};
	\vertex at \coord{150} (B) {};
	\vertex at \coord{30} (C) {};
	\vertex at \coord{90} (D) {};
           \draw [directed,bend left=20] (A) to (D);
           \draw [directed,bend left=-20] (A) to (D);
           \draw [directed,bend left=0] (A) to (D);
	\draw  [directed,bend left=0] (D) to (B);
	\draw  [directed,bend left=0] (D) to (C);
           \draw  [directed,bend left=-70] (A) to (B) ;
           \draw  [directed,bend left=0] (A) to (B) ;
           \draw  [directed,bend left=0] (A) to (C) ;
         \draw  [directed,bend left=70] (A) to (C);
\draw [directed] (D) to (1.5,0);
\draw [directed] (B) to (0.477,-1.422);
\draw [directed] (B) to (0.993,-1.124);
 \draw [directed] (B) to (0.75,-1.3);
\draw [directed] (C) to (0.477,1.422);
\draw [directed] (C) to (0.993,1.124);
 \draw [directed] (C) to (0.75,1.3);
\vcirc at (C) {};
\vbeta at (A) {};
	;
\draw (0,-1.5) node {\small{(c)}};
\\};
\end{tikzpicture}
\caption{Six-loop diagrams for $\gamma_{\phi^n}$ contributing at leading $n$}\label{diagsix}
\end{figure}
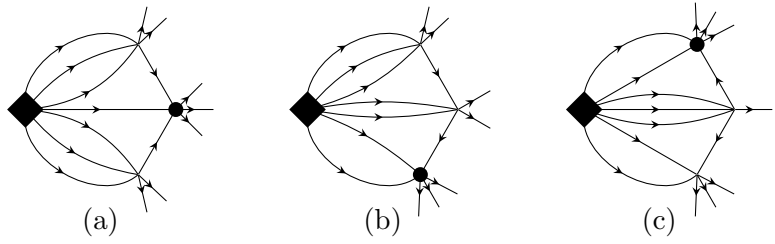

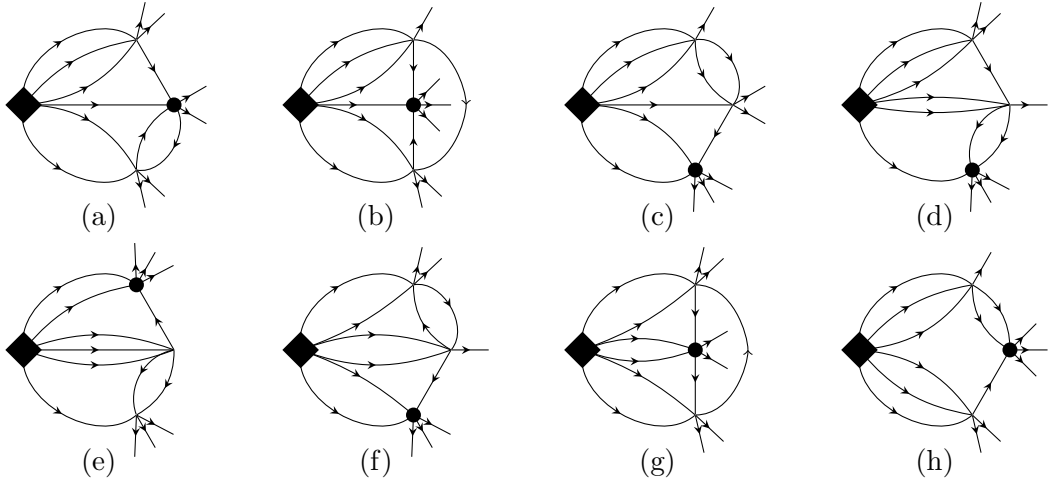
\begin{figure}[ht]
\begin{tikzpicture}
\matrix[column sep=1cm]
{
	\node (vert_cent) {\hspace{-13pt}$\phantom{-}$};
	\vertex at \coord{-90} (A) {};
	\vertex at \coord{150} (B) {};
	\vertex at \coord{30} (C) {};
	\vertex at \coord{90} (D) {};
           \draw [directed] (A) to (D);
	\draw  [directed,bend left=70] (D) to (B);
	\draw  [directed,bend left=0] (C) to (D);
	\draw  [directed,bend left=30] (B) to (D);
         \draw  [directed,bend left=70] (A) to (C);
           \draw  [directed,bend left=-75] (A) to (B) ;
           \draw  [directed,bend left=20] (A) to (C) ;
           \draw  [directed,bend left=30] (A) to (B) ;
           \draw  [directed,bend left=-25] (A) to (C) ;
\draw [directed] (C) to (0.616,1.367);
\draw [directed] (C) to (0.876,1.218);
\draw [directed] (B) to (0.616,-1.367);
\draw [directed] (B) to (0.876,-1.218);
\draw [directed] (D) to (1.43,-0.25);
\draw [directed] (D) to (1.43,0.25);

\vbeta at (A) {};
\vcirc at (D) {};
	;
\draw (0,-1.5) node {\small{(a)}};

&
	\node (vert_cent) {\hspace{-13pt}$\phantom{-}$};
	\vertex at \coord{-90} (A) {};
	\vertex at \coord{150} (B) {};
	\vertex at \coord{30} (C) {};
	\vertex at (0.5,0) (D) {};
	\draw  [directed,bend left=0] (C) to (D);
	\draw  [directed,bend left=0] (A) to (D);
	\draw  [directed,bend left=0] (B) to (D);
         \draw  [directed,bend left=70] (A) to (C);
           \draw  [directed,bend left=-75] (A) to (B) ;
           \draw  [directed,bend left=20] (A) to (C) ;
           \draw  [directed,bend left=30] (A) to (B) ;
           \draw  [directed,bend left=-25] (A) to (C) ;
\draw [->] (C) ..controls (1,.86) and (1.2,.2) .. (1.2,0);
\draw (B) ..controls (1,-.86) and (1.2,-.2) .. (1.2,0);
\draw [directed] (B) to (0.616,-1.367);
\draw [directed] (B) to (0.876,-1.218);
 \draw [directed] (C) to (0.75,1.3);
\draw [directed] (D) to (0.85,-0.35);
\draw [directed] (D) to (0.85,0.35);
\draw [directed] (D) to (1,0);

\vbeta at (A) {};
\vcirc at (D) {};
	;
\draw (0,-1.5) node {\small{(b)}};

&

	\node (vert_cent) {\hspace{-13pt}$\phantom{-}$};
	\vertex at \coord{-90} (A) {};
	\vertex at \coord{150} (B) {};
	\vertex at \coord{30} (C) {};
	\vertex at \coord{90} (D) {};
           \draw [directed] (A) to (D);
	\draw  [directed,bend left=0] (D) to (B);
	\draw  [directed,bend left=70] (C) to (D);
	\draw  [directed,bend left=-30] (C) to (D);
         \draw  [directed,bend left=70] (A) to (C);
           \draw  [directed,bend left=-75] (A) to (B) ;
           \draw  [directed,bend left=20] (A) to (C) ;
           \draw  [directed,bend left=30] (A) to (B) ;
           \draw  [directed,bend left=-25] (A) to (C) ;
 \draw [directed] (C) to (0.75,1.3);
\draw [directed] (B) to (0.477,-1.422);
\draw [directed] (B) to (0.993,-1.124);
 \draw [directed] (B) to (0.75,-1.3);
\draw [directed] (D) to (1.43,-0.25);
\draw [directed] (D) to (1.43,0.25);
\vcirc at (B) {};

\vbeta at (A) {};
	;
\draw (0,-1.5) node {\small{(c)}};

&

	\node (vert_cent) {\hspace{-13pt}$\phantom{-}$};
	\vertex at \coord{-90} (A) {};
	\vertex at \coord{150} (B) {};
	\vertex at \coord{30} (C) {};
	\vertex at \coord{90} (D) {};
           \draw [directed,bend left=10] (A) to (D);
           \draw [directed,bend left=-10] (A) to (D);
	\draw  [reverse directed,bend left=50] (B) to (D);
	\draw  [reverse directed,bend left=-30] (B) to (D);
	\draw  [directed,bend left=0] (C) to (D);
           \draw  [directed,bend left=-70] (A) to (B) ;
           \draw  [directed,bend left=-15] (A) to (C) ;
           \draw  [directed,bend left=20] (A) to (C) ;
         \draw  [directed,bend left=70] (A) to (C);
\draw [directed] (C) to (0.616,1.367);
\draw [directed] (C) to (0.876,1.218);
\draw [directed] (D) to (1.5,0);
\draw [directed] (B) to (0.477,-1.422);
\draw [directed] (B) to (0.993,-1.124);
 \draw [directed] (B) to (0.75,-1.3);

\vbeta at (A) {};
\vcirc at (B) {};
	;
\draw (0,-1.5) node {\small{(d)}};

\\

	\node (vert_cent) {\hspace{-13pt}$\phantom{-}$};
	\vertex at \coord{-90} (A) {};
	\vertex at \coord{150} (B) {};
	\vertex at \coord{30} (C) {};
	\vertex at \coord{90} (D) {};
           \draw [directed,bend left=20] (A) to (D);
           \draw [directed,bend left=-20] (A) to (D);
           \draw [directed,bend left=0] (A) to (D);
	\draw  [reverse directed,bend left=50] (B) to (D);
	\draw  [reverse directed,bend left=-30] (B) to (D);
	\draw  [reverse directed,bend left=0] (C) to (D);
           \draw  [directed,bend left=-70] (A) to (B) ;
           \draw  [directed,bend left=20] (A) to (C) ;
           \draw  [directed,bend left=70] (A) to (C) ;
\draw [directed] (C) to (0.477,1.422);
\draw [directed] (C) to (0.993,1.124);
 \draw [directed] (C) to (0.75,1.3);
\draw [directed] (B) to (0.477,-1.422);
\draw [directed] (B) to (0.993,-1.124);
 \draw [directed] (B) to (0.75,-1.3);

\vbeta at (A) {};
\vcirc at (C) {};
	;
\draw (0,-1.5) node {\small{(e)}};

&

	\node (vert_cent) {\hspace{-13pt}$\phantom{-}$};
	\vertex at \coord{-90} (A) {};
	\vertex at \coord{150} (B) {};
	\vertex at \coord{30} (C) {};
	\vertex at \coord{90} (D) {};
           \draw [directed,bend left=20] (A) to (D);
           \draw [directed,bend left=-20] (A) to (D);
	\draw  [directed,bend left=0] (D) to (B);
	\draw  [directed,bend left=70] (C) to (D);
	\draw  [reverse directed,bend left=-30] (C) to (D);
         \draw  [directed,bend left=70] (A) to (C);
           \draw  [directed,bend left=-70] (A) to (B) ;
           \draw  [directed,bend left=10] (A) to (B) ;
           \draw  [directed,bend left=-10] (A) to (C) ;
\draw [directed] (C) to (0.616,1.367);
\draw [directed] (C) to (0.876,1.218);
\draw [directed] (B) to (0.477,-1.422);
\draw [directed] (B) to (0.993,-1.124);
 \draw [directed] (B) to (0.75,-1.3);
\draw [directed] (D) to (1.5,0);

\vbeta at (A) {};
\vcirc at (B) {};
	;
\draw (0,-1.5) node {\small{(f)}};

&

	\node (vert_cent) {\hspace{-13pt}$\phantom{-}$};
	\vertex at \coord{-90} (A) {};
	\vertex at \coord{150} (B) {};
	\vertex at \coord{30} (C) {};
	\vertex at (0.5,0)  (D) {};
           \draw [directed,bend left=20] (A) to (D);
           \draw [directed,bend left=-20] (A) to (D);
	\draw  [directed,bend left=0] (D) to (B);
	\draw  [directed,bend left =0] (C) to (D);
         \draw  [directed,bend left=70] (A) to (C);
           \draw  [directed,bend left=-70] (A) to (B) ;
           \draw  [directed,bend left=10] (A) to (B) ;
           \draw  [directed,bend left=-10] (A) to (C) ;
\draw [->] (B) ..controls (1,-.86) and (1.2,-.2) .. (1.2,0);
\draw (C) ..controls (1,.86) and (1.2,.2) .. (1.2,0);
\draw [directed] (C) to (0.616,1.367);
\draw [directed] (C) to (0.876,1.218);
\draw [directed] (B) to (0.616,-1.367);
\draw [directed] (B) to (0.876,-1.218);
\draw [directed] (D) to (.93,-0.25);
\draw [directed] (D) to (.93,0.25);

\vbeta at (A) {};
\vcirc at (D) {};
	;
\draw (0,-1.5) node {\small{(g)}};

&

	\node (vert_cent) {\hspace{-13pt}$\phantom{-}$};
	\vertex at \coord{-90} (A) {};
	\vertex at \coord{150} (B) {};
	\vertex at \coord{30} (C) {};
	\vertex at \coord{90} (D) {};
	\draw  [directed,bend left=0] (B) to (D);
	\draw  [directed,bend left=-30] (C) to (D);
	\draw  [directed,bend left=30] (C) to (D);
           \draw  [directed,bend left=-70] (A) to (B) ;
           \draw  [directed,bend left=-20] (A) to (B) ;
           \draw  [directed,bend left=25] (A) to (B) ;
           \draw  [directed,bend left=20] (A) to (C) ;
           \draw  [directed,bend left=-25] (A) to (C) ;
         \draw  [directed,bend left=70] (A) to (C);
 \draw [directed] (C) to (0.75,1.3);
\draw [directed] (B) to (0.616,-1.367);
\draw [directed] (B) to (0.876,-1.218);
\draw [directed] (D) to (1.35,-0.35);
\draw [directed] (D) to (1.35,0.35);
\draw [directed] (D) to (1.5,0);

\vbeta at (A) {};
\vcirc at (D) {};
	;
\draw (0,-1.5) node {\small{(h)}};

\\};
\end{tikzpicture}
\caption{Additional six-loop diagrams for $\gamma_{\phi^n}$ contributing at next-to-leading $n$}\label{diagsixa}
\end{figure}

We now turn to the six-loop calculation. At this loop order we focus from the outset on the contributions leading and subleading in $n$. The leading order six-loop contributions come solely from the diagrams depicted in Fig.~\ref{diagsix} (of course these also produce contributions of lower order in $n$). Once again, the extraction of the poles in $\epsilon$ from these diagrams is described in some detail in the Appendix; the small black circles at the vertices will be explained in that context. 
The next-to-leading contributions six-loop contributions come from the diagrams in Fig.~\ref{diagsix}, together with the additional diagrams depicted in Fig.~\ref{diagsixa}. 

The resulting simple poles for each diagram are tabulated in Table~\ref{6loop}, together with the corresponding symmetry factors. A factor of $\kappa^3$ should also be included for each diagram. For completeness, the full set of pole terms is listed in the Appendix, in Eq.~\eqref{fullpoles}.
\begin{table}
\begin{center}
\begin{tabular}{|c|c|c|} \hline
&&\\
Graph & Symmetry Factor&Simple Pole\\
&&\\
\hline
&&\\
2(a)&$\frac{1}{1728}\frac{n!}{(n-6)!}$&$-\frac{16}{3}$\\
&&\\
\hline
&&\\
2(b)&$\frac{1}{576}\frac{n!}{(n-6)!}$&$\frac{64}{3}$\\
&&\\
\hline
&&\\
2(c)&$\frac{1}{1728}\frac{n!}{(n-6)!}$&$\frac{16}{3}$\\
&&\\
\hline
&&\\
3(a)&$\frac{1}{96}\frac{n!}{(n-5)!}$&$-\frac{2}{3}\pi^2(2\ln2-5)$\\
&&\\
\hline
&&\\
3(b)&$\frac{1}{144}\frac{n!}{(n-5)!}$&$-\frac{16}{9}(\pi^2-12)$\\
&&\\
\hline
&&\\
3(c)&$\frac{1}{288}\frac{n!}{(n-5)!}$&$-\frac{8}{9}(\pi^2-24)$\\
&&\\
\hline
&&\\
3(d)&$\frac{1}{288}\frac{n!}{(n-5)!}$&$\frac{64}{3}$\\
&&\\
\hline
&&\\
3(e)&$\frac{1}{864}\frac{n!}{(n-5)!}$&$\frac{16}{3}$\\
&&\\
\hline
&&\\
3(f)&$\frac{1}{96}\frac{n!}{(n-5)!}$&$\pi^2\left(\frac43\ln2+2\right)$\\
&&\\
\hline
&&\\
3(g)&$\frac{1}{192}\frac{n!}{(n-5)!}$&$\frac{2}{3}\pi^4$\\
&&\\
\hline
&&\\
3(h)&$\frac{1}{864}\frac{n!}{(n-5)!}$&$\frac{8}{9}(\pi^2-6)$\\
&&\\
\hline
\end{tabular}
\caption{\label{6loop}Six-loop results from Figs.~\ref{diagsix}, \ref{diagsixa}}
\end{center}
\end{table}

The contribution to the six-loop anomalous dimension from the diagrams in Figs.~\ref{diagsix}, \ref{diagsixa} is then obtained by adding the products of corresponding symmetry factors and simple poles in Table~\ref{6loop} and multiplying by the usual loop factor of six and a factor $\kappa^3$. The contribution at leading and next-to-leading $n$ is given by
\be
\gamma_{\phi^n}^{(6)} 
= \frac29\kappa^3\left(n^7+\left[-13+\frac{10}{9}\pi^2+\frac{3}{32}\pi^4\right]n^6+\ldots\right),
\ee
in agreement with Eq.~\eqref{Delexp}.

\section{The $O(N)$ case}
In the four-dimensional case the $U(1)$ computation of Ref.~\cite{Bad} was extended to $O(N)$ in Ref.~\cite{sann}. A similar agreement between the semiclassical and perturbative calculations was found, up to three-loop order in perturbation theory. It seems natural to perform a similar extension to $O(N)$ in the case at hand, especially as the group theory and other results developed in Ref.~\cite{sann} may straightforwardly be adapted to $d=3$. Of course this represents a generalisation of the $U(1)$ calculation, since the latter may be recovered as the special case $N=2$; but now we may also wish to consider the limit of large $N$, for example. 

In the $O(N)$ case we have a multiplet of fields $\phi_i$, $i=1\ldots N$, and the Lagrangian is now
\be
{\cal L} =\frac12\pa^{\mu}\phi_i\pa_{\mu}\phi_i+\frac{g^2}{8\times3!}(\phi_i\phi_i)^3.
\ee
As shown in Ref.~\cite{sann}, the fixed-charge operator of charge $\Qbar$ may be taken to be
\be
T_{\Qbar}=T_{i_1i_2\ldots i_{\Qbar}}\phi_{i_1}\phi_{i_2}\ldots \phi_{i_{\Qbar}},
\ee
where $T_{i_1i_2\ldots i_{\Qbar}}$ is symmetric,  and traceless on any pair of indices. The scaling dimension $\Delta_{T_{\Qbar}}$ is expanded in a similar fashion to Eq.~\eqref{scal} as 
\be
\Delta_{T_{\Qbar}}=\Qbar\left(\frac d2-1\right)+\gamma_{T_{\Qbar}}=\sum_{\kappa=-1}\lambda^{\kappa}\Delta_{\kappa}(g\Qbar).
\label{Tscal}
\ee
As in the $U(1)$ case, we initially work in general $d$. The semiclassical computation of $\Delta_{-1}$ and $\Delta_0$ proceeds in a similar manner to the $U(1)$ case, but now the chemical potential $\mu$ is related to the cylinder radius $R$ by
\be
R\mu=\frac{1}{2\sqrt2}\sqrt{1+\sqrt{1+\frac{g^2\Qbar^2}{2\pi^2}}}.
\label{mudefa}
\ee
The computation of the leading contribution is entirely analogous to the $U(1)$ case and is given by
\be
\Delta_{-1}(g \Qbar)=g\Qbar F_{-1}\left(\frac{g^2\Qbar^2}{2\pi^2}\right),
\label{DelmQ}
\ee
where $F_{-1}$ is as defined in Eq.~\eqref{Fdef}. As in the $U(1)$ case, for simplicity we give in Eq.~\eqref{mudefa} the result for $d=3$. The non-leading corrections $\Delta_0$ are once more given by the 
determinant of small fluctuations. There are two modes corresponding to those in the abelian case, with the dispersion relation in Eq.~\eqref{omdef}. In addition there are $\frac{N}{2}-1$ ``Type II'' (non-relativistic)\cite{Niel} Goldstone modes and $\frac{N}{2}-1$ massive states with dispersion relation
\be
\omega_{\pm\pm}(l)=\sqrt{J_l^2+\mu^2}\pm\mu,
\ee
with $J_l$ as defined in Eq.~\eqref{Jdef}. We then find that $\Delta_0$ is given by
\be
\Delta_0(g\Qbar)=\Delta_0^{(a)}(g\Qbar)+\left(\frac{N}{2}-1\right)\Delta_0^{(b)}(g\Qbar),
\label{ONDela}
\ee
where
\begin{align}
\Delta_0^{(a)}(g\Qbar)=&\frac{R}{2}\sum_{l=0}^{\infty}n_l[\omega_+(l)+\omega_-(l)]\nn
\Delta_0^{(b)}(g\Qbar)=&\frac{R}{2}\sum_{l=0}^{\infty}n_l[\omega_{++}(l)+\omega_{--}(l)].
\label{ONDel}
\end{align}
 Here $n_l$ defined in Eq.~\eqref{nldef} is again the multiplicity of the laplacian on the $d$-dimensional sphere, and $\omega_{\pm}$ are defined in \eqref{omdef} but with $R$, $\mu$ now related by Eq.~\eqref{mudefa}.
As before, with a slight abuse of notation, after analytic continuation we replace $\Delta_0^{(a)}(g\Qbar)$,  $\Delta_0^{(b)}(g\Qbar)$ by regularised forms
\begin{align}
\Delta_{0}^{(a)}(g\Qbar)=&\frac14-3(R\mu)^2+\tfrac12\sqrt{8R^2\mu^2-1}+\frac12\sum_{l=1}^{\infty}\sigma^{(a)}(l),\nn
\Delta_{0}^{(b)}(g\Qbar)=&-\frac14-(R\mu)^2+R\mu+\frac12\sum_{l=1}^{\infty}\sigma^{(b)}(l),
\label{Delab}
\end{align}
 where
\begin{align}
\sigma^{(a)}(l)=&(1+2l)R[\omega_+(l)+\omega_-(l)]\nn
&-4l(l+1)-\left(6(R\mu)^2-\frac12\right),\nn
\sigma^{(b)}(l)=&(1+2l)R[\omega_{++}(l)+\omega_{--}(l)]\nn
&-4l(l+1)-\left(2(R\mu)^2+\frac12\right),
\label{sigab}
\end{align}
are defined once again by subtracting positive and zero powers of $l$  in the large-$l$ expansions of Eq.~\eqref{ONDel} so as to give a convergent sum in $d=3$. 
Now expanding $\sigma^{(a)}(l)$, $\sigma^{(b)}(l)$ in powers of $\frac{g^2\Qbar^2}{2\pi^2}$, we find
\begin{align}
\sigma^{(a)}(l)=&C_{2,l}\left(\frac{g^2\Qbar^2}{2\pi^2}\right)^2+C_{3,l}\left(\frac{g^2\Qbar^2}{2\pi^2}\right)^3+\ldots,\nn
\sigma^{(b)}(l)=&\Ctil_{2,l}\left(\frac{g^2\Qbar^2}{2\pi^2}\right)^2+\Ctil_{3,l}\left(\frac{g^2\Qbar^2}{2\pi^2}\right)^3+\ldots,
\label{sigabexp}
\end{align}
where $C_{2,l}$, $C_{3,l}$ were defined in Eq.~\eqref{Cdef}, and
\be
\Ctil_{2,l}=-\frac{1}{128(2l+1)^2},\quad
\Ctil_{3,l}=\frac{16l^2+16l+5}{1024(2l+1)^4}.
\label{Ctil}
\ee

Performing the summations, and combining Eqs.~\eqref{Tscal} (with $d=3$), \eqref{DelmQ}, and \eqref{ONDela}-\eqref{Ctil}, we find the expansion 
\begin{align}
\Delta_{T_{\Qbar}}=&\frac {\Qbar}{2} +\tkappa\left[\frac23(\Qbar^3-3\Qbar^2)+{\cal O}(\Qbar)\right]\nn
&-\tkappa^2\left[4\Qbar^5-\left\{32-\left(4+\frac14N\right)\pi^2\right\}\Qbar^4+{\cal O}(\Qbar^3)\right]\nn
&+\tkappa^3\left[48\Qbar^7+\left\{-624+\left(\frac{136}{3}+4N\right)\pi^2+\frac{1}{12}(52+N)\pi^4\right\}\Qbar^6+{\cal O}(\Qbar^5)\right]+\ldots,
\label{DelexpQ}
\end{align}
where $\tkappa=\left(\frac{g}{8\pi}\right)^2$. We note that the $U(1)$ result in the previous section may be obtained by setting $N=2$ and making the substitution $g^2=\tfrac16\lambda^2$.
\begin{table}
\begin{center}
\begin{tabular}{|c|c|} \hline
&\\
Graph & Result\\
&\\
\hline
&\\
$A$ & $\frac{\Qbar(\Qbar-1)(\Qbar-2)(\Qbar-3)(\Qbar-4)}{4}I_4$\\
&\\
\hline
& \\
$B$ & $\frac98\Qbar(\Qbar-1)(\Qbar-2)(\Qbar-3)\frac{1}{18}(16+N)I_{4bbb}$\\
&\\
\hline
&\\
$C$ & $\frac{\Qbar(\Qbar-1)(\Qbar-2)(\Qbar-3)}{2}I_4$\\
&\\
\hline
&\\
$D$ & $\frac92\Qbar(\Qbar-1)(\Qbar-2)I_4$\\
&\\
\hline
&\\
$E$ & $\frac92\Qbar(\Qbar-1)\Ytil$\\

&\\ 
\hline
\end{tabular}
\caption{\label{4loopa}Four-loop results for $O(N)$ case}
\end{center}
\end{table}

\begin{table}
\begin{center}
\begin{tabular}{|c|c|c|} \hline
&&\\
Graph & Symmetry Factor&Simple Pole\\
&&\\
\hline
&&\\
2(a)&$\frac{1}{8}\frac{\Qbar!}{(\Qbar-6)!}$&$-\frac{16}{3}$\\
&&\\
\hline
&&\\
2(b)&$\frac{3}{8}\frac{\Qbar!}{(\Qbar-6)!}$&$\frac{64}{3}$\\
&&\\
\hline
&&\\
2(c)&$\frac{1}{8}\frac{\Qbar!}{(\Qbar-6)!}$&$\frac{16}{3}$\\
&&\\
\hline
&&\\
3(a)&$\frac{9}{4}\frac{\Qbar!}{(\Qbar-5)!}\frac{1}{432}(384+24N)$&$-\frac{2}{3}\pi^2(2\ln2-5)$\\
&&\\
\hline
&&\\
3(b)&$\frac{3}{2}\frac{\Qbar!}{(\Qbar-5)!}$&$-\frac{16}{9}(\pi^2-12)$\\
&&\\
\hline
&&\\
3(c)&$\frac{3}{4}\frac{\Qbar!}{(\Qbar-5)!}$&$-\frac{8}{9}(\pi^2-24)$\\
&&\\
\hline
&&\\
3(d)&$\frac{3}{4}\frac{\Qbar!}{(\Qbar-5)!}$&$\frac{64}{3}$\\
&&\\
\hline
&&\\
3(e)&$\frac{1}{4}\frac{\Qbar!}{(\Qbar-5)!}$&$\frac{16}{3}$\\
&&\\
\hline
&&\\
3(f)&$\frac{9}{4}\frac{\Qbar!}{(\Qbar-5)!}\frac{1}{432}(384+24N)$&$\pi^2\left(\frac43\ln2+2\right)$\\
&&\\
\hline
&&\\
3(g)&$\frac{9}{8}\frac{\Qbar!}{(\Qbar-5)!}\frac{1}{432}(416+8N)$&$\frac{2}{3}\pi^4$\\
&&\\
\hline
&&\\
3(h)&$\frac{1}{4}\frac{\Qbar!}{(\Qbar-5)!}$&$\frac{8}{9}(\pi^2-6)$\\
&&\\
\hline
\end{tabular}
\caption{\label{6loopa}Six Loop Results for $O(N)$ case}
\end{center}
\end{table}

The contributions from individual diagrams for this case are shown in Tables~\ref{4loopa} and \ref{6loopa}. Factors of $\tkappa^2$ at four loops and $\tkappa^3$ at six loops are implicit. As mentioned before, the $U(1)$ results may be recovered by setting $N=2$ and making the substitution $g^2=\tfrac16\lambda^2$. Once again, after adding the diagrammatic contributions and including a loop factor of 4 and 6 respectively, the leading and subleading four and six loop contributions  agree with the semiclassical result in Eq.~\eqref{DelexpQ}. It is noteworthy that the $N$ dependence in Eq.~\eqref{DelexpQ} involves purely powers of $\pi^2$; and this feature in fact appears to persist to higher orders. It would be interesting to be able to associate this with a generic topological property of the relevant Feynman diagrams.

\section{Large $g\Qbar$}
In the $U(1)$ case, the result for the anomalous dimension may be expanded for large $\lambda n$ and compared with the effective theory for the gapless Goldstone
 mode corresponding to $\omega_-$. In the $O(N)$ case, we can do an analogous expansion for large $g\Qbar$. Following Ref.~\cite{Bad}, we obtain
\begin{align}
\Delta_{T_{\Qbar}}=\tbar^{\frac32}\left[c_{3/2}+c_{1/2}\tbar^{-1}+c_{-1/2}\tbar^{-2}+\ldots\right]\nn
+\left[d_0+d_{-1}\tbar^{-1}+\ldots\right].
\label{numfit}
\end{align}
with $\tbar=\frac{\sqrt2g\Qbar}{\pi}$ and
\be
c_i=\ctil_i+\left(\frac{N}{2}-1\right)\cbar_i,\quad d_i=\ctil_i+\left(\frac{N}{2}-1\right)\dbar_i,
\ee
where
\begin{align}
\ctil_{3/2}\approx&\frac{\pi}{6\sqrt2g}-0.0653+{\cal O}\left(\frac{\sqrt2g}{\pi}\right),\nn
\ctil_{1/2}\approx&\frac{\pi}{2\sqrt2g}+0.2088+{\cal O}\left(\frac{\sqrt2g}{\pi}\right),\nn
\ctil_{-1/2}\approx&-\frac{\pi}{4\sqrt2g}-0.2627+{\cal O}\left(\frac{\sqrt2g}{\pi}\right),\nn
\dtil_0\approx&-0.0937255,\nn
\dtil_{-1}\approx&0.096+{\cal O}\left(\frac{\sqrt2g}{\pi}\right).
\label{expvals}
\end{align}
and 
\begin{align}
\cbar_{3/2}\approx&-0.010417,\nn
\cbar_{1/2}\approx&0.052083,\nn
\cbar_{-1/2}\approx&-0.096875,\nn
\dbar_0\approx&\dbar_1\approx0,
\label{cvals}
\end{align}
The leading-order contributions in Eq.~\eqref{expvals} follow straightforwardly from expanding $\Delta_{-1}$ in Eq.~\eqref{DelmQ} for large $g\Qbar $  using Eq.~\eqref{Fdef}. The next-to-leading order numbers, in Eqs.~\eqref{expvals} and \eqref{cvals}, derive from a numerical fit to $\Delta_0$ as given by Eq.~\eqref{ONDel}, following the procedure explained in Ref.~\cite{Bad} and in more detail in Ref.~\cite{Bad2}. 
 
The values in Eqs.~\eqref{expvals} were essentially given already in Ref.~\cite{Bad}, after making allowance for the change from $\lambda n$ to $g\Qbar$. The numerical coefficients $\cbar_i$ in Eq.~\eqref{cvals} are therefore the only new features of the $O(N)$ case at large $g\Qbar$. We note the intriguing fact that $\cbar_{3/2}=5\cbar_{1/2}$. This fact and indeed the values of the remaining $\cbar_i$ may be explained quite simply. It is convenient to consider an expansion of $\Delta_0$ in powers of $v=R\mu$, rather than $g\Qbar$; of course in view of Eq.~\eqref{mudefa}, large $g\Qbar$ implies large $R\mu$. We find from redoing the numerical matching 
\be
\Delta_0=a_3v^3+a_2v^2+a_1v+a_0+\frac{a_{-1}}{v}+\frac{a_{-2}}{v^2}++\frac{a_{-3}}{v^3}+\ldots,
\label{vser}
\ee
with
\be
a_i=\atil_i+\left(\frac{N}{2}-1\right)\abar_i,
\ee
where
\begin{align}
\atil_3\approx&-4.1812,\nn
\atil_2\approx&0,\nn
\atil_1\approx&1.6192,\nn
\atil_0\approx&-0.093725,\nn
\atil_{-1}\approx&-0.09334,\nn
\atil_{-2}\approx&0.006051,\nn
\atil_{-3}\approx&-0.003911,
\label{atvals}
\end{align}
and 
\begin{align}
\abar_3\approx-2\abar_1\approx&-\frac23,\nn
\abar_2\approx\abar_0\approx\abar_{-2}\approx\abar_{-4}\approx&0,\nn
\abar_{-1}\approx&-\frac{1}{30},\nn
\abar_{-3}\approx&-0.0031746,\nn
\abar_{-5}\approx&-0.0011992. 
\label{avals}
\end{align}
It is easy to see by expanding $R\mu$ in Eq.~\eqref{mudef} that the coefficients $\dtil_i$ and $\dbar_i$ in Eq.~\eqref{numfit} depend only on the even powers of $R\mu$, $\atil_{2j}$ and $\abar_{2j}$, respectively, and indeed we see from Eq.~\eqref{avals} that the $\abar_{2j}$ and the $\dbar_i$ all vanish. It is also easy to check that the $\ctil_i$ and $\cbar_i$ coefficients in Eqs.~\eqref{expvals} and \eqref{cvals} are derived from the odd coefficients $\atil_{2j+1}$ and $\abar_{2j+1}$.
For instance we have 
\be
\cbar_{3/2}=\frac{1}{64}\abar_3, \quad \cbar_{1/2}=\frac{1}{64}(3\abar_3+16\abar_1)\quad \cbar_{-1/2}=\frac{1}{128}(9\abar_3+32\abar_1+512\abar_{-1}),
\ee
(with similar relations for $\ctil_i$ and $\atil_i$)
which are easily verified using the values in Eqs.~\eqref{expvals}, \eqref{cvals}, \eqref{atvals} and \eqref{avals}. The previously-noted relation $\cbar_{3/2}=5\cbar_{1/2}$ is seen to follow from the relation $\abar_3=-2\abar_1$ in Eq.~\eqref{avals}.
The values of the coefficients of $\abar_i$ for negative $i$ may now be understood as follows. Once again separating $\Delta_0$ as in Eq.~\eqref{ONDel}, we find from the analytic expansion of Eq.~\eqref{Delab} for large $v$ 
\begin{align}
\Delta_0^{(a)}=&\sum_{l=0}^{\infty}\Bigl\{-2l(l+1)-3v^2+\frac14\nn
&+\sqrt2(1+2l)\left(v+\frac14\Jtil_l-\frac{1}{32}\frac{2-3\Jtil_l^2}{v}+\frac{1}{128}\frac{\Jtil_l^3-\Jtil_l}{v^2}\right)\Bigr\},\nn
\Delta_0^{(b)}=&\sum_{l=0}^{\infty}\Bigl\{-2l(l+1)-v^2-\frac14\nn
&+(1+2l)\left(v+\frac{\Jtil_l^2}{2v}-\frac{\Jtil_l^4}{8v^3}+\frac{\Jtil_l^6}{16v^5}+\ldots\right)\Bigr\},
\label{newser}
\end{align} 
where 
\be
\Jtil_l^2=R^2J_l^2=l(l+1)
\ee
with $J_l^2$ as in Eq.~\eqref{Jdef}, but with $d=3$. We would now be able to reproduce Eq.~\eqref{vser} with Eqs.~\eqref{atvals}, \eqref{avals}, if we could perform the summations over $l$.  However, it turns out that we can only make progress on this in the case of $\Delta_0^{(b)}$. Its two crucial properties appear to be the following: it has an expansion in powers of $\frac{\Jtil_l^2}{v^2}$, with a leading term $v$, and, as we explained earlier, the leading positive/zero powers in the large-$l$ expansion have been subtracted in Eq.~\eqref{Delab} (as they also were for $\Delta_0^{(a)}$, of course). An immediate consequence is that there are no negative even powers of $v$ in $\Delta_0^{(b)}$ in Eq.~\eqref{newser}, implying the vanishing of the $\abar_{2j}$ for $j$ negative. The summations in Eq.~\eqref{newser} are all {\it a priori} infinite. Nevertheless, it turns out that we can obtain regularised results for those in $\Delta_0^{(b)}$ corresponding to odd powers of $v$. If we write
\be
\zeta(s)=\sum_{l=1}^{\infty}l^{-s}
\label{zdef}
\ee
then we can define
\begin{align}
\sum_{l=0}^{\infty}l^n=&\zeta(-n)=(-1)^n\frac{B_{n+1}}{n+1},\quad (n>0),\nn
\sum_{l=0}^{\infty}l^0=&\zeta(0)+1=B_1+1,
\label{lsums}
\end{align}
where $B_n$ are the Bernoulli numbers.  In the second sum in Eq.~\eqref{lsums}, we have accounted for the fact that the series in Eq.~\eqref{newser},  Eq.~\eqref{zdef}, start at $l=0$, $l=1$, respectively; of course this makes no difference in the first sum. We obtain the following expressions for the coefficients:
\begin{align}
\abar_1=&\phantom{-}[-B_2+B_1+1]=\frac13,\nn
\abar_{-1}=&-\frac12\left[-\frac12B_{4}+\frac12B_2\right]=\phantom{-}\frac14\left(\frac{1}{30}-\frac16\right)=-\frac{1}{30},\nn
\abar_{-3}=&\phantom{-}\frac18\left[\frac13B_6+B_4\right]=-\frac{1}{8}\left[-\frac13.\frac{1}{42}+\frac{1}{30}\right]\nn
=&-\frac{1}{315}\approx-0.0031746,\nn
\abar_{-5}=&-\frac{1}{16}\left[\frac14B_8+\frac32B_6+\frac14B_4\right]
=\phantom{-}\frac{1}{64}\left[\frac{1}{30}-6\frac{1}{42}+\frac{1}{30}\right]\nn
=&-\frac{1}{840}\approx-0.0011905.
\end{align}
recalling that $B_n=0$ for $n$ odd, except for $n=1$.  Comparing with Eq.~\eqref{avals}, we find surprisingly good agreement. 
Turning now to $\abar_0$ and $\abar_2$,  the cancellation of leading powers of $l$ in Eq.~\eqref{Delab} appears to guarantee the vanishing of these coefficients as observed in Eq.~\eqref{avals}, even though the $\zeta$-function sums defined by Eq.~\eqref{lsums} do not give vanishing results for the $v^2$ and $v^0$ terms in Eq.~\eqref{newser}.
 We have checked that for other functions sharing the crucial properties mentioned above, we similarly obtain $\abar_i=0$ for $i\le2$ and even; and the $\zeta$-function sums correctly give $\abar_i$ for $i\le1$ and odd.  However, $\abar_3$ remains a problem. There is no $v^3$ term in Eq.~\eqref{newser} to match the one in Eq.~\eqref{vser}; though if one approximates the original sum over $l$ in Eq.~\eqref{ONDel} by an integral, one easily sees the emergence of a $v^3$ term, with indeed the correct coefficient.

On the other hand, although the definition of $\Delta_0^{(a)}$  in Eq.~\eqref{Delab} correctly subtracts the leading $l^2$ and $l^0$ terms, the large-$v$ expansion in Eq.~\eqref{newser} does not have the other crucial property mentioned above. Consequently it contains odd powers of $\Jtil_l$ (associated with negative even powers of $v$)
 and hence factors of $\sqrt{l(l+1)}$ which cannot be summed using Eq.~\eqref{lsums}. Furthermore, the $\zeta$-function sums for the odd powers of $v$ fail to agree with the results obtained in Eq.~\eqref{atvals}. It is then no surprise that $\atil_2$ and $\atil_0$ in Eq.~\eqref{atvals} fail to vanish, as might otherwise have been expected from our experience with $\Delta_0^{(b)}$.

 Nevertheless, we have succeeded in obtaining exact expressions for the ``new'' coefficients in the large $R\mu$, and consequently large $g \Qbar$, expansions in the $O(N)$ case (i.e. those coefficients which are not already present in the $U(1)$ case); albeit we have no rigorous explanation for the values of $\abar_3=-\frac23$, $\abar_2=\abar_0=0$.

\section{Conclusions}

Neutron stars, and high density quark matter can both be described in terms of a superfluid effective field theory for a Goldstone boson field\cite{son}\cite{bcmr}. As explained in Ref.~\cite{Bad},   relevant issues may also be addressed in terms of the relativistic  theory of a complex scalar field $\phi$  with $\left(\frac{\lambda}{3!}\right)^2(\phi^*\phi)^3$ interactions,   in $d= 3-\epsilon$ dimensions.  This theory has a conformal fixed point (for small $\epsilon$) at 
\be \left(\frac{\lambda}{3!}\right)^2=  \frac{3}{7}\epsilon. \ee

In this paper we have extended the calculation of the anomalous dimension of the operator $\phi^n$ embarked upon in Ref.~\cite{Bad} from two loops to four and six loops. We continue to find agreement between the straightforward perturbative (in $\lambda^2$) calculation and the results of a semiclassical calculation, along the lines explained in Ref.~\cite{Bad2}.  This agreement interpolates between large and small $\lambda n$. 

We performed similar calculations for an $O(N)$ theory with $(\phi^i\phi^i)^3$  interactions, which includes the  $U(1)$ case described above as the special case $N=2$. Here both semiclassical and perturbative approaches were pursued in Ref~\cite{sann}, for 
$(\phi^i\phi^i)^2$  theory in $d=4-\epsilon$, which similarly has a conformal fixed point with a coupling constant of $O(\epsilon)$. It turns out to be quite straightforward to adapt these calculations to the $d=3-\epsilon$ case, and once again we find that the perturbative and semiclassical approaches interpolate seamlessly into one another. The conformal fixed point is crucial to the semiclassical discussion. In the large-$N$ limit, the coupling must be rescaled, and the conformal fixed point is changed so the discussion would require modification\footnote{The fixed point structure in $d=3$ at large $N$ is explored in Ref.~\cite{hughster}}; we do not pursue this issue here.  Finally in the $O(N)$ case we have shown how to compute exactly the $N$-dependent parts of the coefficients in the large charge expansion.

\section*{Acknowledgements} 
We thank John Gracey and Hugh Osborn for conversations and Andrei Kataev and Diego Rodriguez-Garcia for correspondence. 
DRTJ thanks the Leverhulme Trust for the award of an Emeritus Fellowship. This research was supported by the
Leverhulme Trust, STFC and by the University of Liverpool.

\appendix

\section{Full diagram results}
In this appendix we explain in some detail how we have derived our perturbative results. We start with a pedagogical description of the four-loop calculation; the techniques are well-known to high-loop experts but maybe not to the wider community and not in the three-dimensional context. 

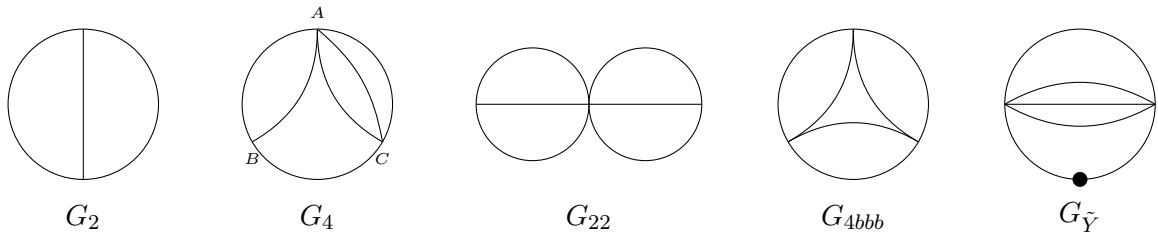
\begin{figure}[ht]
\begin{tikzpicture}
\matrix[column sep=1cm]
{
\node (vert_cent) {\hspace{-13pt}$\phantom{-}$};
	\draw circle[radius=1cm](0,0);
	\node[fill,circle,inner sep=0pt,minimum size=0pt] at \coord{0} (A) {};
	\vertex at \coord{180} (B) {};
\draw(A) to (B);
\draw (0,-1.5) node {\small{$G_2$}};
&

	\node (vert_cent) {\hspace{-13pt}$\phantom{-}$};
	\draw circle[radius=1cm](0,0);
	\node[fill,circle,inner sep=0pt,minimum size=0pt] at \coord{0} (A) {};
	\vertex at \coord{120} (B) {};
	\vertex at \coord{-120}(C) {};
	\draw[bend right] (A) to (B);
\draw[bend left=20] (A) to (B);
\draw[bend left] (A) to (C);
\draw (0,1) node[above] {\tiny{$A$}};
\draw (B) node[below] {\tiny{$C$}};
\draw (C) node[below] {\tiny{$B$}};
\draw (0,-1.5) node {\small{$G_4$}};
&

	\draw (-0.75,0) circle[radius=0.75cm];
	\draw (0.75,0) circle[radius=0.75cm];
\draw (-1.5,0) -- (1.5,0);
\draw (0,-1.5) node {\small{$G_{22}$}};

&

	\node (vert_cent) {\hspace{-13pt}$\phantom{-}$};
	\draw circle[radius=1cm](0,0);
	\node[fill,circle,inner sep=0pt,minimum size=0pt] at \coord{0} (A) {};
	\vertex at \coord{120} (B) {};
	\vertex at \coord{-120}(C) {};
	\draw[bend right] (A) to (B);
	\draw[bend right] (B) to (C);
\draw[bend left] (A) to (C);
\draw (0,-1.5) node {\small{$G_{4bbb}$}};

&

	\draw (0,0) circle[radius=1cm];
\draw[bend left] (-1,0) to (1,0);
\draw[bend right] (-1,0) to (1,0);
\draw (-1,0) -- (1,0);
\vcirc at (0,-1) {};
\draw (0,-1.5) node {\small{$G_{\tilde Y}$}};

\\};
\end{tikzpicture}
\caption{Two- and four-loop momentum integrals}\label{fourmom}
\end{figure}

We define the result of the generic one-loop integral by the ``$G$-function'' $G(a,b)$\cite{kataev}, so that
\be
G(a,b)=\int d^dk\frac{p^{2(a+b-\frac{d}{2})}}{k^{2a}(p-k)^{2b}}=\frac{\Gamma\left(a+b-\frac{d}{2}\right)\Gamma\left(\frac{d}{2}-a\right)\Gamma\left(\frac{d}{2}-b\right)}{(4\pi)^{\frac{d}{2}}\Gamma(a)\Gamma(b)\Gamma(d-a-b)}.
\ee
The first divergence appears in the two-loop integral $G_2$ depicted in Fig.~\ref{fourmom}, and given by 
\be
G_2=G_1G\left(2-\tfrac12d,1\right)
\ee
where for convenience we denote the basic one-loop bubble by $G_1=G(1,1)$. The pole term is given by
\be
I_2=\Khat[G_2]=\frac{1}{64\pi^2}\frac{2}{\epsilon},
\ee
where $\Khat$ denotes the operation of extracting the divergent part. Note that our convention in this paper is that $G$ denotes the full momentum integral and $I$ the corresponding local counterterm (after subtracting subdivergences where necessary; see later). We are using minimal subtraction, so the counterterm is purely divergent. In a slight misuse of notation, $G$ will often be used to  refer both to the graph and to the corresponding Feynman integral.

The four-loop momentum integrals are also depicted in Fig.~\ref{fourmom}. The Feynman graphs corresponding to all diagrams considered in this paper (two, four and six loop) are logarithmically divergent and consequently the extraction of the counterterms may be simplified using ``infra-red (IR) rearrangement''.\footnote{This technique was used in very early ``multi-loop" calculations\cite{Jones},\cite{Caswell}.} This involves judiciously setting selected external momenta to zero, leaving a single momentum entering at one vertex and leaving at another, in order to obtain a more tractable integral.
\ It will be useful to focus on $G_4$ for our pedagogical introduction. For convenience we have labelled the vertices of $G_4$ in Fig.~\ref{fourmom} by $A$, $B$, $C$. We first consider the case where a momentum enters at $A$ and leaves at $B$. The basic momentum integral is given by
\be
G_4=  G_1G_2G\left(2-\tfrac12d,4-d\right).
\ee
There is also a divergent two-loop subgraph, with a divergence $I_2$, which needs to be subtracted to obtain a local result.  We obtain
\be
I_4=\Khat\Rbar[G_4]=\Khat\left[G_2\left\{G_1G\left(2-\tfrac12d,4-d\right)-I_2\right\}\right]=\frac{1}{(64\pi^2)^2}\frac{1}{\epsilon^2}(-2+4\epsilon).
\ee
The process of correctly subtracting the subdivergences is here denoted $\Rbar$.  For more details see Ref.~\cite{klein} where the procedure is well explained (with reference to the four-dimensional case). In general there may be several distinct ways of implementing the IR rearrangement. Any IR rearrangement which avoids the introduction of spurious IR divergences will give the same result for the final counterterm, after making the appropriate subtractions. In the case of $I_4$, for instance, we may also consider the case where a momentum enters at $C$ and leaves at $B$. The basic momentum integral is then given by
\be
G'_4=G_1G_2G\left(5-\tfrac32d,1\right),
\ee
and we now have 
\be
I_4=\Khat\Rbar[G'_4]=\Khat\left[G_2\left\{G_1G\left(5-\tfrac32d,1\right)-I_2\right\}\right]=\frac{1}{(64\pi^2)^2}\frac{1}{\epsilon^2}(-2+4\epsilon).
\ee
As emphasised earlier, the same result is obtained for the counterterm $I_4$. In general, in the process of IR rearrangement, the same entry and exit points must be used for the subtracted diagrams as for the original. For a different IR rearrangement of a given diagram, the pole terms for the original diagram and the subtracted diagrams will typically be individually different, but will combine to give the same total counterterm. In the current case, the subtractions are the same for the two IR rearrangements. The same overall result is nevertheless obtained for the pole term since both $G_4$ and $G_4'$ also have the same poles, though of course differ in their finite parts.

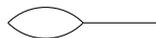
\begin{figure}[ht]
\center\begin{tikzpicture}
\draw[bend left=45] (-1,0) to (0,0);
\draw[bend right=45] (-1,0) to (0,0);
\draw (0,0) -- (1,0);
\end{tikzpicture}
\caption{Basic IR-divergent structure}\label{IRdiv}
\end{figure}

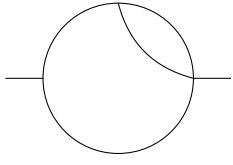
\begin{figure}
\center\begin{tikzpicture}
	\node (vert_cent) {\hspace{-13pt}$\phantom{-}$};
	\draw circle[radius=1cm](0,0);
	\node[fill,circle,inner sep=0pt,minimum size=0pt] at \coord{0} (A) {};
	\vertex at \coord{90} (B) {};
	\vertex at \coord{-90}(C) {};
	\draw[bend right] (A) to (B);
\draw (-1.5,0) -- (-1,0);
\draw (1.5,0) -- (1,0);
\end{tikzpicture}
\caption{Two-loop IR-divergent integral}\label{IRtwo}
\end{figure}

There is a final  possible IR rearrangement, where the momentum enters at $A$ and leaves at $C$. This requires a more careful treatment. In four dimensions one is familiar with the basic IR divergence from a double propagator; in the current case of  three dimensions, the basic IR divergence is a double propagator followed by a single one, as shown in Fig.~\ref{IRdiv}. 
This structure leads to an effective propagator 
\be
G_1\frac{1}{(p^2)^{3-\frac12d}}
\label{effprop}
\ee
where the IR divergence in three dimensions is clearly revealed. It appears in this third IR rearrangement which consequently leads to a spurious IR divergence. We may avoid this spurious divergence by using the $\Rbar^*$ procedure, which augments the $\Rbar$ procedure with a subtraction for the IR divergences
\cite{chet}. 
We start by considering the basic two-loop IR-divergent diagram in Fig.~\ref{IRtwo}.
This is given by
\be
G_1G\left(1,3-\tfrac12d\right)\sim\frac{1}{64\pi^2}\left(-\frac{2}{\epsilon}\right);
\ee
the divergence coming from $\Gamma(d-3)$ where the positive sign for $d$ signals the infra-red nature of the divergence. The IR subtraction for this simple, single IR divergence is consequently to replace Eq.~\eqref{effprop} by
\be
\left(G_1\frac{1}{(p^2)^{3-\frac12d}}+\frac{2}{64\pi^2\epsilon}\delta(p)\right).
\label{newprop}
\ee  
The IR divergence is cancelled in Fig.~\ref{IRtwo} when \eqref{newprop} is used to replace \eqref{effprop}. Just as we saw earlier for the case of IR rearrangement combined with the standard $\Rbar$ procedure, the same process must be applied to the subtracted diagrams. Expressed diagrammatically, denoting the IR-subtracted propagator of Eq.~\eqref{newprop} by a line with a box, we have
\begin{align}
I_4=&\Khat\left[\tikz[baseline=(vert_cent.base)]{
\node (vert_cent) {\hspace{-13pt}$\phantom{-}$};
	\draw circle[radius=1cm](0,0);
	\node[fill,circle,inner sep=0pt,minimum size=0pt] at \coord{0} (A) {};
	\vertex at \coord{180} (B) {};
	\draw[bend right] (A) to (B);
\draw[bend left] (A) to (B);
\draw[bend right] (A) to (B);
\vsq at (-1,0) {};
\vcirc at (0,1) {};
\vcirc at (0,-1) {};
}
-I_2
\tikz[baseline=(vert_cent.base)]{
\node (vert_cent) {\hspace{-13pt}$\phantom{-}$};
	\draw circle[radius=1cm](0,0);
	\node[fill,circle,inner sep=0pt,minimum size=0pt] at \coord{0} (A) {};
	\vertex at \coord{180} (B) {};
\vsq at (-1,0) {};
\vcirc at (1,0) {};
}\right]\nn
=&\Khat\left[\tikz[baseline=(vert_cent.base)]{
\node (vert_cent) {\hspace{-13pt}$\phantom{-}$};
	\draw circle[radius=1cm](0,0);
	\node[fill,circle,inner sep=0pt,minimum size=0pt] at \coord{0} (A) {};
	\vertex at \coord{180} (B) {};
	\vertex at \coord{-90}(C) {};
	\draw[bend left=45] (A) to (B);
\draw  (A) -- (B);
\draw[bend left] (A) to (C);
\vcirc at (0,1) {};
\vcirc at (0,-1) {};
}
+\frac{2}{64\pi^2\epsilon}\tikz[baseline=(vert_cent.base)]{
\node (vert_cent) {\hspace{-13pt}$\phantom{-}$};
	\draw circle[radius=1cm](0,0);
	\node[fill,circle,inner sep=0pt,minimum size=0pt] at \coord{0} (A) {};
	\vertex at \coord{180} (B) {};
\draw (A) -- (B);
\vcirc at (0,1) {};
\vcirc at (0,-1) {};
}
-I_2\tikz[baseline=(vert_cent.base)]{
\node (vert_cent) {\hspace{-13pt}$\phantom{-}$};
	\draw circle[radius=1cm](0,0);
\draw (0,1) -- (0,-1);
\vcirc at (0,1) {};
	\node[fill,circle,inner sep=0pt,minimum size=0pt] at \coord{0} (A) {};
}
-\frac{2}{64\pi^2\epsilon}I_2\right]\nn
=&\Khat[G_1G_2G(3-\tfrac12d,3-d)+\frac{2}{64\pi^2\epsilon}G_2-\frac{2}{64\pi^2\epsilon}I_2].
\end{align}
Here we denote the momentum entrance and exit points by a small black circle. A diagram with a single black circle has coincident momentum entrance and exit points and vanishes in dimensional regularisation.
Once again, the same result is obtained for $I_4$.

The corresponding pole terms for the remaining four-loop diagrams in Fig.~\ref{fourmom} are given by
\begin{align}
I_{22}=&\Khat\Rbar[G_{22}]=\Khat\Rbar[G_2^2]=\Khat[G_2(G_2-2I_2)]=\frac{1}{(64\pi^2)^2}\frac{1}{\epsilon^2}(-4),\nn
I_{4bbb}=&\Khat[G_{4bbb}]=\Khat\left[G_1^3G\left(4-d,2-\tfrac12d\right)\right]=\frac{1}{(64\pi^2)^2}\frac{1}{\epsilon}\pi^2,\nn
{\tilde Y}=&\Khat[G_1^2G\left(2,2-\tfrac12d\right)G\left(4-d,2-\tfrac12d\right)]=\frac{1}{(64\pi^2)^2}\left(-\frac{2}{\epsilon}\right),
\label{fourpole}
\end{align}
where $G_{22}$ and $G_{4bbb}$ are implicitly defined in  terms of $G$-functions.

 We now turn to the six-loop computation, for which the diagrams are shown in Figs.~\ref{diagsix} and \ref{diagsixa}. These are again logarithmically divergent. Once again we use IR rearrangement, so that we retain just a single incoming and outgoing momentum; in all our examples, this momentum may be thought of as entering at the $\phi^n$ vertex (i.e. the lozenge) since it turns out that this ensures wherever possible that the result may readily be expressed in terms of $G$-functions. The momentum leaves at the vertex marked by the small black circle. As observed earlier in the case of $I_4$,  the choice of momentum entrance and exit  is not unique; but once made, must also be used for the subtracted diagrams. We have made the choice of momentum exit so as (for simplicity) where possible to avoid introducing infra-red divergences, either in the six-loop diagram itself or in its subtractions ; even though such IR divergences may be accommodated using the $\Rbar^*$ procedure. 
In the case of diagrams with a structure such as Fig.~\ref{IRdiv}, the potential IR divergence may be avoided by choosing the central vertex in Fig.~\ref{IRdiv} as the exit for the momentum. It will be observed that this has been done in Fig.~\ref{diagsix}(b) and Figs.~\ref{diagsixa}(f), (h). Less obviously, the choice of momentum exit in Fig.~\ref{diagsixa}(b) has been made to avoid an IR divergence in the two-loop subtracted diagram.  The process of infra-red rearrangement also reduces the number of independent integrals; for instance, with the choice of momentum exit indicated, Figs. \ref{diagsixa}(e) and \ref{diagsix}(c) correspond to the same integral, despite the structure Fig.~\ref{IRdiv} being reversed in the latter diagram. Figs.~\ref{diagsixa}(d) and \ref{diagsix}(b) look different, since the lower single loop is on different sides of the momentum exit point. However, either of the alternative IR rearrangements using one of the other $(\phi\phi^*)^3$ vertices as exit would make the pair of diagrams look identical, and therefore would demonstrate that Figs.~\ref{diagsixa}(d) and \ref{diagsix}(b) produce the same counterterm after subtractions. Of course we could have used one of those alternative IR rearrangements, but at the expense of being obliged to use the (slightly more complicated) $\Rbar^*$ procedure.
In all the cases mentioned so far, IR rearrangement leads to graphs which may easily be evaluated in terms of $G$-functions. For those where the simple $\Rbar$ procedure is sufficient, we find
\begin{align}
I_{2a}=&\Khat[G_2^2G(4-d,1)G\left(5-\tfrac32d,4-d\right)-2I_{2}G_4-I_{22}G_2],\nn
I_{2b}=&I_{3d}=\Khat[G_1G_4G\left(2-\tfrac12d,7-2d\right)-I_{2}G_4-I_4G_2],\nn
I_{3a}=&\Khat[G_1^2G_2G(4-d,1)G\left(5-\tfrac32d,4-d\right)-I_{2}G_{4bbb}],\nn
I_{3c}=&\Khat[G_1^2G_2G\left(5-\tfrac32d,1\right)G\left(2-\tfrac12d,7-2d\right)-I_{2}G_4-I_4G_2],\nn
I_{3f}=&\Khat[G_1^4G\left(4-d,2-\tfrac12d\right)G\left(2-\tfrac12d,7-2d\right)-I_{4bbb}G_2],\nn
I_{3h}=&\Khat[G_2^2G_1G\left(5-\tfrac32d,4-d\right)-I_{2}G_4-I_2G_4'-I_{22}G_2],\nn
\end{align}
Here we use $I_{2a}$ (for instance) to represent the counterterm resulting (after subtraction of subdivergences) from the Feynman integral $G_{2a}$ corresponding to Fig.~\ref{diagsix}(a), just as $I_4$ results from $G_4$. We shall give a complete list of explicit expressions for the pole terms later, after discussing the range of general procedures required for the different classes of diagram. We emphasise that, as explained earlier, the expression for $I_{3d}$ in terms of diagram plus subtractions for the IR rearrangement shown in Fig.~\ref{diagsixa}(d) would be different from that shown explicitly for $I_{2b}$, but the final total would be the same.
In just one instance, Fig.~\ref{diagsix}(c) (or equivalently Fig.~\ref{diagsixa}(e)), the process of IR rearrangement  inevitably introduces infra-red divergences due to the presence of two IR-divergent structures of the form Fig.~\ref{IRdiv}. One of these must then be dealt with using the $\Rbar^*$ operation\cite{chet} explained earlier in the context of $I_4$. We find
\begin{align}
I_{2c}=&\Khat\Bigl[\tikz[baseline=(vert_cent.base)]{
\node (vert_cent) {\hspace{-13pt}$\phantom{-}$};
	\vertex at \coord{-90} (A) {};
	\vertex at \coord{180} (B) {};
	\vertex at \coord{30} (C) {};
	\vertex at \coord{90} (D) {};
           \draw [bend left=20] (A) to (D);
           \draw [bend left=-20] (A) to (D);
           \draw [bend left=0] (A) to (D);
	\draw  [bend left=-60] (B) to (D);
	\draw  [bend left=0] (C) to (D);
           \draw  [bend left=60] (B) to (A) ;
           \draw  [bend left=0] (A) to (C) ;
           \draw  [bend left=70] (A) to (C) ;
\vcirc at (C) {};
\vsq at (B) {};
\vcirc at (A) {};
}
-I_2\tikz[baseline=(vert_cent.base)]{
\draw (-0.75,0) circle[radius=0.75cm];
	\draw (0.75,0) circle[radius=0.75cm];
\draw (0,0) -- (1.5,0);
\vcirc at (0,0) {};
\vcirc at (1.5,0) {};
\vsq at (-1.5,0) {};
}-I_4\tikz[baseline=(vert_cent.base)]{
\draw (-0.75,0) circle[radius=0.75cm];
\draw (-1.5,0) -- (0,0);
\vcirc at (-1.5,0) {};
\vcirc at (0,0) {};
}
-I_4\tikz[baseline=(vert_cent.base)]{
\draw (-0.75,0) circle[radius=0.75cm];
\vsq at (-1.5,0) {};
\vcirc at (0,0) {};
}\Bigr]
\nn
=&\Khat\Bigl[\tikz[baseline=(vert_cent.base)]{
	\node (vert_cent) {\hspace{-13pt}$\phantom{-}$};
	\vertex at \coord{-90} (A) {};
	\vertex at \coord{150} (B) {};
	\vertex at \coord{30} (C) {};
	\vertex at \coord{90} (D) {};
           \draw [bend left=20] (A) to (D);
           \draw [bend left=-20] (A) to (D);
           \draw [bend left=0] (A) to (D);
	\draw  [bend left=0] (D) to (B);
	\draw  [bend left=0] (C) to (D);
           \draw  [bend left=-70] (A) to (B) ;
           \draw  [bend left=0] (A) to (B) ;
           \draw  [bend left=0] (A) to (C) ;
         \draw  [bend left=70] (A) to (C);
\vcirc at (C) {};
\vcirc at (A) {};
}+\frac{2}{64\pi^2\epsilon}\tikz[baseline=(vert_cent.base)]{
	\node (vert_cent) {\hspace{-13pt}$\phantom{-}$};
	\vertex at \coord{-90} (A) {};
	\vertex at \coord{150} (B) {};
	\vertex at \coord{30} (C) {};
	\vertex at \coord{90} (D) {};
           \draw [bend left=20] (A) to (D);
           \draw [bend left=-20] (A) to (D);
           \draw [bend left=0] (A) to (D);
	\draw  [bend left=0] (C) to (D);
           \draw  [bend left=0] (A) to (C) ;
         \draw  [bend left=70] (A) to (C);
\vcirc at (C) {};
\vcirc at (A) {};
}\nn
&-I_2\tikz[baseline=(vert_cent.base)]{
\draw (-0.75,0) circle[radius=0.75cm];
	\draw (0.75,0) circle[radius=0.75cm];
\draw (-1.5,0) -- (1.5,0);
\vcirc at (0,0) {};
\vcirc at (1.5,0) {};
}-\frac{2}{64\pi^2\epsilon}I_2\tikz[baseline=(vert_cent.base)]{
\draw (-0.75,0) circle[radius=0.75cm];
\draw (-1.5,0) -- (0,0);
\vcirc at (0,0) {};
\vcirc at (-1.5,0) {};
}-I_4\tikz[baseline=(vert_cent.base)]{
\draw (-0.75,0) circle[radius=0.75cm];
\draw (-1.5,0) -- (0,0);
\vcirc at (-1.5,0) {};
\vcirc at (0,0) {};
}\nn
&
-I_4\tikz[baseline=(vert_cent.base)]{
\draw (-0.75,0) circle[radius=0.75cm];
\draw (-1.5,0) -- (0,0);
\vcirc at (0,0) {};
}-I_4\frac{2}{64\pi^2\epsilon}
\Bigr]\nn
=&\Khat\Bigl[\left\{G_1^2G_2G\left(3-\tfrac12d,3-d\right)G\left(2-\tfrac12d,7-2d\right)+\frac{2}{64\pi^2\epsilon}G_4\right\}\nn
&-\frac{2}{64\pi^2\epsilon}I_2G_2-I_4G_2-I_4\frac{2}{64\pi^2\epsilon}\Bigr].
\end{align}
Once again, diagrams with a single black circle have coincident momentum entrances and exits and vanish in dimensional regularisation.

In a couple of cases, namely Figs.~ \ref{diagsixa}(b) and \ref{diagsixa}(g), there is no IR rearrangement which leads simply to an expression in terms of $G$-functions, and we need to use an identity\cite{gracey2} derived using the ``integration by parts'' trick\cite{tkachov}\cite{chet2}, which enables us to simplify integrals of the form shown in Fig.~\ref{diagspec} which occur as substructures in these graphs. In this diagram, $\alpha_i$, $i=1\ldots5$, represent the weights of the corresponding propagators.
\begin{figure}
\center\begin{tikzpicture}
{
  \node (vert_cent) {\hspace{-13pt}$\phantom{-}$};
\draw (-3,0)--(-2,0);
 \draw(-2,0)--(0,2);
\draw(-2,0)--(0,-2);
\draw(2,0)--(0,2);
\draw(2,0)--(0,-2);
    \draw (0,2)--(0,-2);
    \draw (2,0)--(3,0);    
\draw (-1,-1) node[below]{$\small{\alpha_4}$};       
\draw (-1,1) node[above]{$\alpha_1$};  
\draw (1,1) node[above]{$\alpha_2$};  
\draw (1,-1) node[below]{$\alpha_3$};  
\draw (0.2,0.2) node[below]{$\alpha_5$};  
        }
\end{tikzpicture}
\caption{Diagram} \label{diagspec}
\end{figure}
This identity is given here in diagrammatic form.
\begin{align}
(d-\alpha_1-\alpha_4-2\alpha_5)\tikz[baseline=(vert_cent.base)]{
  \node (vert_cent) {\hspace{-13pt}$\phantom{-}$};
\draw (-1.5,0)--(-1,0);
 \draw(-1,0)--(0,1);
\draw(-1,0)--(0,-1);
\draw(1,0)--(0,1);
\draw(1,0)--(0,-1);
    \draw (0,1)--(0,-1);
    \draw (1,0)--(1.5,0);      
        }
=&\alpha_1\tikz[baseline=(vert_cent.base)]{
  \node (vert_cent) {\hspace{-13pt}$\phantom{-}$};
\draw (-1.5,0)--(-1,0);
 \draw(-1,0)--(0,1);
\draw(-1,0)--(0,-1);
\draw(1,0)--(0,1);
\draw(1,0)--(0,-1);
    \draw (0,1)--(0,-1);
    \draw (1,0)--(1.5,0);   
\draw (-0.5,0.5) node[above]{$+$};  
\draw (0.2,0.2) node[below]{$-$};     
        }
-\alpha_1\tikz[baseline=(vert_cent.base)]{
  \node (vert_cent) {\hspace{-13pt}$\phantom{-}$};
\draw (-1.5,0)--(-1,0);
 \draw(-1,0)--(0,1);
\draw(-1,0)--(0,-1);
\draw(1,0)--(0,1);
\draw(1,0)--(0,-1);
    \draw (0,1)--(0,-1);
    \draw (1,0)--(1.5,0);   
\draw (-0.5,0.5) node[above]{$+$};  
\draw (0.5,0.5) node[above]{$-$};     
        }\nn
&+\alpha_4\tikz[baseline=(vert_cent.base)]{
  \node (vert_cent) {\hspace{-13pt}$\phantom{-}$};
\draw (-1.5,0)--(-1,0);
 \draw(-1,0)--(0,1);
\draw(-1,0)--(0,-1);
\draw(1,0)--(0,1);
\draw(1,0)--(0,-1);
    \draw (0,1)--(0,-1);
    \draw (1,0)--(1.5,0);   
\draw (-0.5,-0.5) node[below]{$+$};  
\draw (0.2,0.2) node[below]{$-$};     
        }
-\alpha_4\tikz[baseline=(vert_cent.base)]{
  \node (vert_cent) {\hspace{-13pt}$\phantom{-}$};
\draw (-1.5,0)--(-1,0);
 \draw(-1,0)--(0,1);
\draw(-1,0)--(0,-1);
\draw(1,0)--(0,1);
\draw(1,0)--(0,-1);
    \draw (0,1)--(0,-1);
    \draw (1,0)--(1.5,0);   
\draw (-0.5,-0.5) node[below]{$+$};  
\draw (0.5,-0.5) node[below]{$-$};     
        }
\label{johnid}
\end{align}
Here a $\pm$ indicates that the weight has been increased/decreased by 1, relative to the weights in Fig.~\ref{diagspec}.
\begin{figure}[ht]
\center\begin{tikzpicture}
  \node (vert_cent) {\hspace{-13pt}$\phantom{-}$};
\draw (-3,0)--(-2,0);
 \draw(-2,0)--(0,2);
\draw(-2,0)--(0,-2);
\draw(2,0)--(0,2);
\draw(2,0)--(0,-2);
    \draw (0,2)--(0,-2);
    \draw (2,0)--(3,0);    
\draw  [bend left=-45] (-2,0) to (0,-2.5);
\draw  [bend left=45] (2,0) to (0,-2.5);
\draw (-1,-1) node[below]{$\alpha_2$};       
\draw (-1,1) node[above]{$\alpha_1$};  
\draw (0,-2.5) node[below]{$\alpha_3$}; 
\draw (1,1) node[above]{$1$};  
\draw (1,-1) node[below]{$1$};  
\draw (0.2,0.2) node[below]{$1$};  
\end{tikzpicture}
\caption{Diagram} \label{diagmodel}
\end{figure}
 After performing simple one- and two-loop integrals, Figs.~\ref{diagsixa}(b) and (g) lead to integrals of the diagrammatic form shown in Fig.~\ref{diagmodel}, which will be denoted $H(\alpha_1,\alpha_2,\alpha_3)$. It is clear that this is a special case of a diagram formed by adding an extra line joining the left and right vertices of Fig.~\ref{diagspec}. The identity in Eq.~\eqref{johnid} may therefore be applied. In this special case, in each diagram on the right-hand side of Eq.~\eqref{johnid} a propagator is cancelled, contracting two vertices and leaving a diagram which may easily be evaluated in terms of $G$-functions. We obtain diagrammatically 
\begin{align}
(d-\alpha_1-\alpha_2-2)H(\alpha_1,\alpha_2,\alpha_3)=&\alpha_1\tikz[baseline=(vert_cent.base)]{
  \node (vert_cent) {\hspace{-13pt}$\phantom{-}$};
 \draw (-0.4,0)--(0.1,0);
    \draw (0.7,0) ++(0:0.6cm and 0.4cm) arc (0:180:0.6cm and 0.4cm) node(n1) {}
             (0.7,0) ++(180:0.6cm and 0.4cm) arc (180:360:0.6cm and 0.4cm) node(n2){};
\draw (1.3,0) ++(180:1.2cm and 0.8cm) arc (180:360:1.2cm and 0.8cm) node(n3) {};
 \node (vert_cent) {\hspace{-13pt}$\phantom{-}$};
    \draw (1.9,0) ++(0:0.6cm and 0.4cm) arc (0:180:0.6cm and 0.4cm) node(n1) {}
             (1.9,0) ++(180:0.6cm and 0.4cm) arc (180:360:0.6cm and 0.4cm) node(n2){};
             \draw (2.5,0)--(3.0,0); 
\draw (.6,.3) node[above]{\tiny{$\alpha_1+1$}}; 
\draw (.6,-0.4) node[above]{\tiny{$\alpha_2$}}; 
\draw (1.3,-0.9) node[above]{\tiny{$\alpha_3$}};
 \draw (1.9,.3) node[above]{\tiny{$1$}}; 
\draw (1.9,-0.4) node[above]{\tiny{$1$}}; 
        }
-\alpha_1\tikz[baseline=(vert_cent.base)]{
  \node (vert_cent) {\hspace{-13pt}$\phantom{-}$};
 \draw (-0.4,0)--(0.1,0);
    \draw (0.7,0) ++(0:0.6cm and 0.4cm) arc (0:180:0.6cm and 0.4cm) node(n1) {}
             (0.7,0) ++(180:0.6cm and 0.4cm) arc (180:360:0.6cm and 0.4cm) node(n2){};
\draw (1.3,0) ++(180:1.2cm and 0.8cm) arc (180:360:1.2cm and 0.8cm) node(n3) {};
 \node (vert_cent) {\hspace{-13pt}$\phantom{-}$};
    \draw (1.9,0) ++(0:0.6cm and 0.4cm) arc (0:180:0.6cm and 0.4cm) node(n1) {}
             (1.9,0) ++(180:0.6cm and 0.4cm) arc (180:360:0.6cm and 0.4cm) node(n2){};
             \draw (1.3,0)--(1.3,0.5); 
\draw (.6,.3) node[above]{\tiny{$\alpha_1+1$}}; 
\draw (.6,-0.4) node[above]{\tiny{$\alpha_3$}}; 
\draw (1.3,-0.9) node[above]{\tiny{$\alpha_2$}};
 \draw (1.9,.3) node[above]{\tiny{$1$}}; 
\draw (1.9,-0.4) node[above]{\tiny{$1$}}; 
        }\nn
&+\alpha_2\tikz[baseline=(vert_cent.base)]{
  \node (vert_cent) {\hspace{-13pt}$\phantom{-}$};
 \draw (-0.4,0)--(0.1,0);
    \draw (0.7,0) ++(0:0.6cm and 0.4cm) arc (0:180:0.6cm and 0.4cm) node(n1) {}
             (0.7,0) ++(180:0.6cm and 0.4cm) arc (180:360:0.6cm and 0.4cm) node(n2){};
\draw (1.3,0) ++(180:1.2cm and 0.8cm) arc (180:360:1.2cm and 0.8cm) node(n3) {};
 \node (vert_cent) {\hspace{-13pt}$\phantom{-}$};
    \draw (1.9,0) ++(0:0.6cm and 0.4cm) arc (0:180:0.6cm and 0.4cm) node(n1) {}
             (1.9,0) ++(180:0.6cm and 0.4cm) arc (180:360:0.6cm and 0.4cm) node(n2){};
             \draw (2.5,0)--(3.0,0); 
\draw (.6,.3) node[above]{\tiny{$\alpha_1$}}; 
\draw (.6,-0.4) node[above]{\tiny{$\alpha_2+1$}}; 
\draw (1.3,-0.9) node[above]{\tiny{$\alpha_3$}};
 \draw (1.9,.3) node[above]{\tiny{$1$}}; 
\draw (1.9,-0.4) node[above]{\tiny{$1$}}; 
        }
-\alpha_2\tikz[baseline=(vert_cent.base)]{
  \node (vert_cent) {\hspace{-13pt}$\phantom{-}$};
 \draw (-0.4,0)--(0.1,0);
    \draw (0.7,0) ++(0:0.6cm and 0.4cm) arc (0:180:0.6cm and 0.4cm) node(n1) {}
             (0.7,0) ++(180:0.6cm and 0.4cm) arc (180:360:0.6cm and 0.4cm) node(n2){};
\draw (1.3,0) ++(180:1.2cm and 0.8cm) arc (180:360:1.2cm and 0.8cm) node(n3) {};
 \node (vert_cent) {\hspace{-13pt}$\phantom{-}$};
    \draw (1.9,0) ++(0:0.6cm and 0.4cm) arc (0:180:0.6cm and 0.4cm) node(n1) {}
             (1.9,0) ++(180:0.6cm and 0.4cm) arc (180:360:0.6cm and 0.4cm) node(n2){};
             \draw (1.3,0)--(1.3,0.5); 
\draw (.6,.3) node[above]{\tiny{$\alpha_2+1$}}; 
\draw (.6,-0.4) node[above]{\tiny{$\alpha_3$}}; 
\draw (1.3,-0.9) node[above]{\tiny{$\alpha_1$}};
 \draw (1.9,.3) node[above]{\tiny{$1$}}; 
\draw (1.9,-0.4) node[above]{\tiny{$1$}}; 
        },
\end{align}
or, in terms of $G$-functions
\begin{align}
H(\alpha_1,\alpha_2,\alpha_3)=&\frac{G(1,1)}{d-\alpha_1-\alpha_2-2}\Bigl[-\alpha_1G(\alpha_1+1,\alpha_2)G(\alpha_1+\alpha_2+3-d,\alpha_3)\nn&+\alpha_1G(\alpha_1+1,\alpha_3)G\left(\alpha_1+\alpha_3+1-\tfrac12d,\alpha_2+2-\tfrac12d\right)\nn
&-\alpha_2G(\alpha_2+1,\alpha_1)G(\alpha_1+\alpha_2+3-d,\alpha_3)\nn&+\alpha_2G(\alpha_2+1,\alpha_3)G\left(\alpha_2+\alpha_3+1-\tfrac12d,\alpha_1+2-\tfrac12d\right)\Bigr].
\end{align}
The diagrams Fig.~\ref{diagsixa}(b) and (g) may therefore be evaluated. Their subtractions are perfectly standard, and we obtain for the pole terms
\begin{align}
I_{3b}=&\Khat\Rbar[G_{3b}]=\Khat\left[G_2G_1H\left(3-d,2-\tfrac12d,1\right)-I_2G_4-I_4G_2
\right],\nn
I_{3g}=&\Khat\Rbar[G_{3g}]=\Khat\left[G_1^3H\left(2-\tfrac12d,2-\tfrac12d,2-\tfrac12d\right)\right].
\end{align}

Finally we can give the full set of pole terms. The final results for the pole terms for the diagrams in Fig.~\ref{diagsix} are

\begin{align}
(64\pi^2)^3I_{2a}=&\frac83\frac{1}{\epsilon^3}(1-2\epsilon-2\epsilon^2),\nn
(64\pi^2)^3I_{2b}=&\frac43\frac{1}{\epsilon^3}(1-6\epsilon+16\epsilon^2),\nn
(64\pi^2)^3I_{2c}=&\frac83\frac{1}{\epsilon^3}(1-4\epsilon+2\epsilon^2);
\end{align}
and the results for the diagrams in  Fig.~\ref{diagsixa} are
\begin{align}
(64\pi^2)^3I_{3a}=&-\frac23\frac{1}{\epsilon^2}\pi^2[1+(2\ln2-5)\epsilon],\nn
(64\pi^2)^3I_{3b}=&\frac43\frac{1}{\epsilon^3}\left[1-6\epsilon-\frac43(\pi^2-12)\epsilon^2\right],\nn
(64\pi^2)^3I_{3c}=&\frac43\frac{1}{\epsilon^3}\left[1-6\epsilon-\frac23(\pi^2-24)\epsilon^2\right],\nn
I_{3d}=&I_{2b},\nn 
I_{3e}=&I_{2c},\nn
(64\pi^2)^3I_{3f}=&\frac{1}{\epsilon^2}\pi^2\left[-\frac43+\left(\frac43\ln2+2\right)\epsilon\right],\nn
(64\pi^2)^3I_{3g}=&\frac23\frac{1}{\epsilon}\pi^4,\nn
(64\pi^2)^3I_{3h}=&\frac83\frac{1}{\epsilon^3}\left[1-2\epsilon+\frac13(\pi^2-6)\epsilon^2\right].
\label{fullpoles}
\end{align}
We notice that it is only primitive diagrams (which have no divergent subdiagrams and therefore only simple pole divergences) which give simple poles with a single order of transcendentality; namely $I_{4bbb}$ in Eq.~\eqref{fourpole} and $I_{3g}$ in Eq.~\eqref{fullpoles}. These produce simple poles with the maximal order of transcendentality for the corresponding loop order: $\pi^2$ at four loops and $\pi^4$ at six loops. We note that the diagrams in Fig.~\ref{diagsixa} are all topologically identical to diagrams contributing to the six-loop $\beta$-function for the $O(N)$ scalar theory, which was computed in Ref.~\cite{hager}, and consequently the corresponding counterterms were computed in that paper; but unfortunately results for individual diagrams are not listed explicitly there.

\end{document}